\pdfoutput=1 
\documentclass[aps,prx,twocolumn,superscriptaddress, longbibliography, 10pt]{revtex4-2}
\usepackage{amsmath}    
\usepackage{graphicx}
\UseRawInputEncoding
\usepackage{booktabs}
\usepackage{xcolor,soul}
\usepackage{amssymb}
\usepackage{mathdots}
\definecolor{darkblue}{HTML}{3771C8}
\usepackage[colorlinks=true,
    pdfborder={0 0 0},
    linkcolor=darkblue,
    citecolor = darkblue, 
    urlcolor = darkblue
]{hyperref}
\usepackage[capitalize,nameinlink]{cleveref}
\usepackage[normalem]{ulem}
\usepackage{braket}
\usepackage{xr}
\makeatletter

\renewcommand{\selectlanguage}[1]{}
\begin{document}
\title{Quantum-enabled continuous microwave-to-optics frequency conversion}

\author{Han Zhao}
    \email[These authors contributed equally.]{}
    \affiliation{Moore Laboratory of Engineering, California Institute of Technology, Pasadena, California 91125}
    \affiliation{Institute for Quantum Information and Matter, California Institute of Technology, Pasadena, California 91125}
\author{William David Chen}
    \email[These authors contributed equally.]{}
    \affiliation{Moore Laboratory of Engineering, California Institute of Technology, Pasadena, California 91125}
    \affiliation{Institute for Quantum Information and Matter, California Institute of Technology, Pasadena, California 91125}
\author{Abhishek Kejriwal}

    \affiliation{Moore Laboratory of Engineering, California Institute of Technology, Pasadena, California 91125}
    \affiliation{Institute for Quantum Information and Matter, California Institute of Technology, Pasadena, California 91125}

\author{Mohammad Mirhosseini}
    \email[mohmir@caltech.com; http://qubit.caltech.edu]{}
    \affiliation{Moore Laboratory of Engineering, California Institute of Technology, Pasadena, California 91125}
    \affiliation{Institute for Quantum Information and Matter, California Institute of Technology, Pasadena, California 91125}

\date{\today} 

\begin{abstract} 
A quantum interface between microwave and optical photons is essential for entangling remote superconducting quantum processors. To preserve fragile quantum states, a transducer must operate efficiently while generating less than one photon of noise referred to its input. Here, we present a platform that meets these criteria, utilizing a combination of electrostatic and optomechanical interactions in devices made entirely from crystalline silicon. This platform's small mechanical dissipation and low optical absorption enable ground-state radiative cooling, resulting in quantum-enabled operation with a continuous laser drive. Under the optimal settings for high efficiency (low noise), we measure an external efficiency of $2.2\%$ ($0.47\%$) and an input-referred added noise of $0.94$ ($0.58$)  in microwave-to-optics conversion. We quantify the transducer throughput using the efficiency-bandwidth product, finding it exceeds previous demonstrations with similar noise performance by approximately two orders of magnitude, thereby paving a practical path to interconnecting remote superconducting qubits.
\end{abstract}
\maketitle

\section*{Introduction} 
Photons, the quantum particles of electromagnetic waves, are ideal carriers of information due to their high propagation speed and minimal interaction with the environment. A multitude of technologies, including global internet infrastructure, take advantage of optical links for remote communication. With progress in our abilities to compute and sense quantum signals, a growing range of quantum information systems also require long-distance and multi-node connectivity \cite{Kimble2008Jun}. Quantum communication, much like its classical counterpart, often relies on photons \cite{Gisin2007Mar}. Interfaces between qubits and photons are thus actively pursued for a variety of physical platforms \cite{Kurizki2015Mar}. Notably, superconducting qubits -- electrical circuits developed for quantum computing -- require microwave-optical interfaces for room-temperature connectivity \cite{Han2021Aug}.

To effectively transmit quantum states, microwave-to-optical transducers have to possess multiple characteristics. A transducer must exhibit high efficiency, ideally nearing $100\%$, for deterministic quantum state transfer \cite{10.1038/s41467-022-34373-8}. A more relaxed set of conditions can be defined for non-deterministic communication, which can be achieved with no fundamental threshold for efficiency so far as the noise is sufficiently low. While the exact noise threshold for operation in this so-called \emph{quantum-enabled} regime \cite{Sahu2022Mar} is protocol-specific, an input noise of below one photon is a widely used condition \cite{Zeuthen2020May, Kumar2023Mar}. In this regime, a practical metric for the transducer throughput is set by the product of efficiency and the state-conversion bandwidth, which bounds the output photon flux and subsequently the remote entanglement generation rate \cite{Krastanov2021Jul,10.1103/physrevapplied.18.054061}.

Achieving both high throughput and low-noise transduction has been challenging despite steady progress in the past \cite{Fan2018Aug,McKenna2020Dec,Holzgrafe2020Dec,Fu2021May,Xu2021Jul,Sahu2022Mar,Sahu2023May,Bartholomew2020Jun,Rochman2023Mar,Kumar2023Mar,Borowka2024Jan,Andrews2014Apr,Arnold2020Sep,Mirhosseini2020Dec,brubaker_optomechanical_2022,Jiang2023Oct,Meesala2023Dec,Weaver2024Feb,Meesala2024Feb}. Initial approaches focused on maximizing conversion efficiency through strong direct interactions between microwaves and light \cite{Fan2018Aug,McKenna2020Dec,Holzgrafe2020Dec,Fu2021May,Xu2021Jul,Sahu2022Mar,Sahu2023May} or via intermediaries like atoms \cite{Bartholomew2020Jun,Rochman2023Mar,Kumar2023Mar,Borowka2024Jan}, magnons \cite{Zhu2020Oct}, or mechanical oscillators \cite{Andrews2014Apr,Arnold2020Sep,Mirhosseini2020Dec,brubaker_optomechanical_2022,Jiang2023Oct,Meesala2023Dec,Weaver2024Feb,Meesala2024Feb}. However, it was soon found that the required strong optical and electrical pumps -- which are essential to bridge the vast frequency gap between microwaves and light -- result in thermal noise. To combat this effect, experiments have turned to pulsed operations where strong pumps are supplied only for short bursts to minimize heating. This approach has led to successful demonstrations in the quantum-enabled regime, including state conversion from qubits as well as microwave-optics quantum entanglement \cite{Mirhosseini2020Dec,Sahu2023May,Meesala2024Feb,Meesala2023Dec}.  Despite these successful demonstrations, the prolonged idling intervals in pulsed operation have significantly reduced the output photon flux, undermining the feasibility of networking operations with two or more nodes.

\begin{figure*}[t!]
\centering
\includegraphics[width=\linewidth]{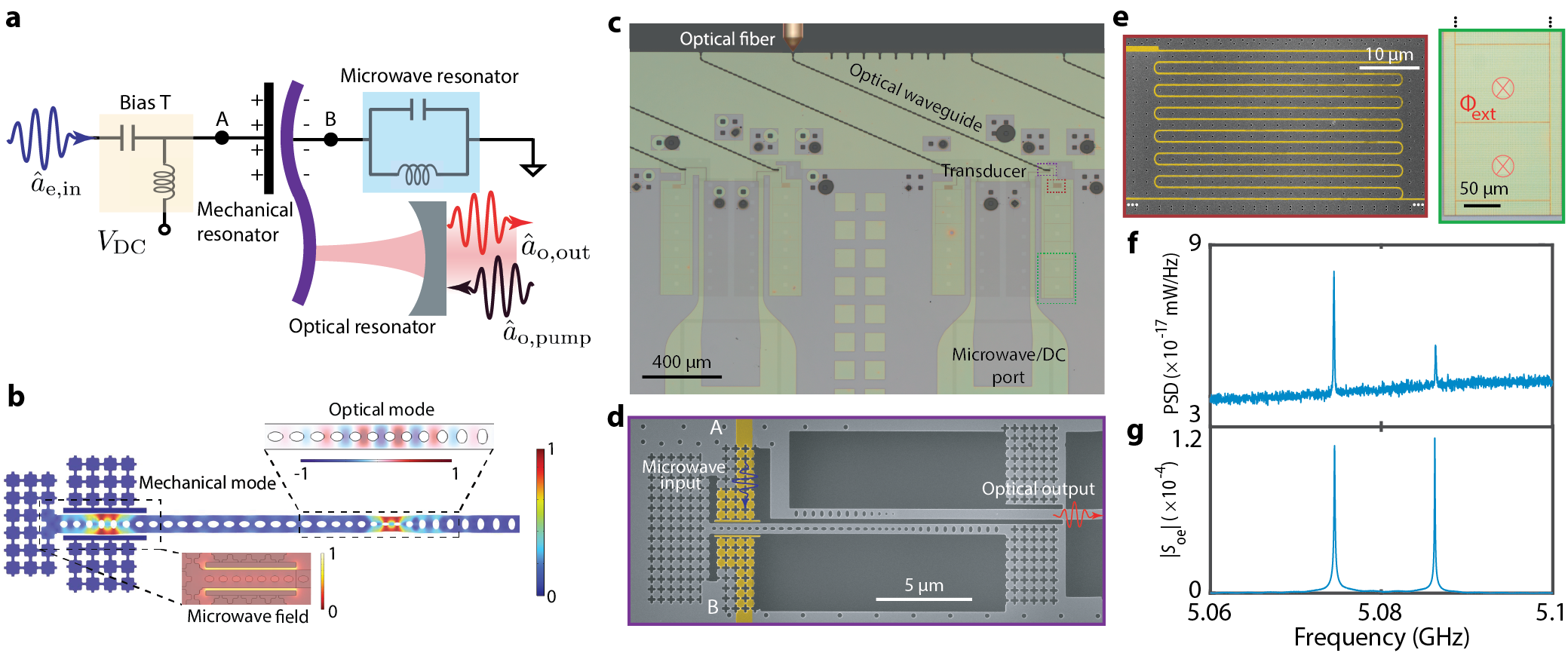}
\caption{\textbf{Electro-optomechanical transduction.} \textbf{a} Schematic diagram of an electro-optomechanical transducer. \textbf{b} Simulated displacement profile of the mechanical mode, showcasing its extended energy distribution. The insets show the microwave and optical field profiles. \textbf{c} Optical microscope image of a fabricated chip containing an array of transducer devices. The optical coupling port is laterally offset by $\sim$1.2 mm to avoid stray light illumination of the microwave resonator. \textbf{d} Scanning electron microscope image of a transducer. The nanobeam is surrounded by acoustic shields with a bandgap at the mechanical frequency. The capacitor electrodes are shown in false color. The letters designate the connection to the rest of the system as shown in panel (a). Optical access is provided by evanescent coupling to a nearby optical waveguide. \textbf{e}  A close-up view of the microwave resonator meander section (scanning electron microscope image, left), which serves to increase the impedance of the microwave resonator for an enhanced electromechanical coupling rate ($g_{\text{em}}\propto \sqrt{Z}$). The square loop array section (optical microscope image, right) enables tuning of the microwave resonance frequency via the dependence of the kinetic inductance on an external magnetic field. \textbf{f} Power spectral density (PSD) of mechanical thermal motion measured via optomechanical scattering of a red-detuned pump. \textbf{g} Coherent microwave-to-optics scattering spectrum at $V_{\text{DC}}= 20$ V.}
\label{figure 1}
\end{figure*}

Here, we introduce an integrated electro-optomechanical platform capable of quantum-enabled performance under continuous operation. In this platform, we eliminate the need for electrical pumping by employing a form of electromechanical interaction that is based on electrostatic fields \cite{Bozkurt2023Sep}. This process's material-agnostic nature allows us to construct the mechanical oscillator entirely from single-crystal silicon, a material with exceptionally low acoustic loss and low optical absorption \cite{Meenehan2014Jul,Meenehan2015Oct,MacCabe2020Nov}. These inherent properties significantly increase cooperativities and reduce heating, enabling ground-state radiative cooling of the mechanical oscillator. Consequently, the device achieves microwave-to-optical conversion with an input-referred added noise well below 1 photon. By operating continuously, our experiment harvests the entire efficiency-bandwidth product, unlike pulsed schemes where the pulse duty cycle reduces the throughput. As a result, we achieve a range of throughputs ($0.47-1.9 \text{ kHz}$) that exceeds previous demonstrations with similar noise performance by about two orders of magnitude. Our experiment thus establishes a new standard for microwave-to-optical quantum frequency conversion and offers a viable path toward the near-term realization of remote entanglement between distant superconducting qubits.

\begin{figure*}[htbp]
\centering
\includegraphics[width=\linewidth]{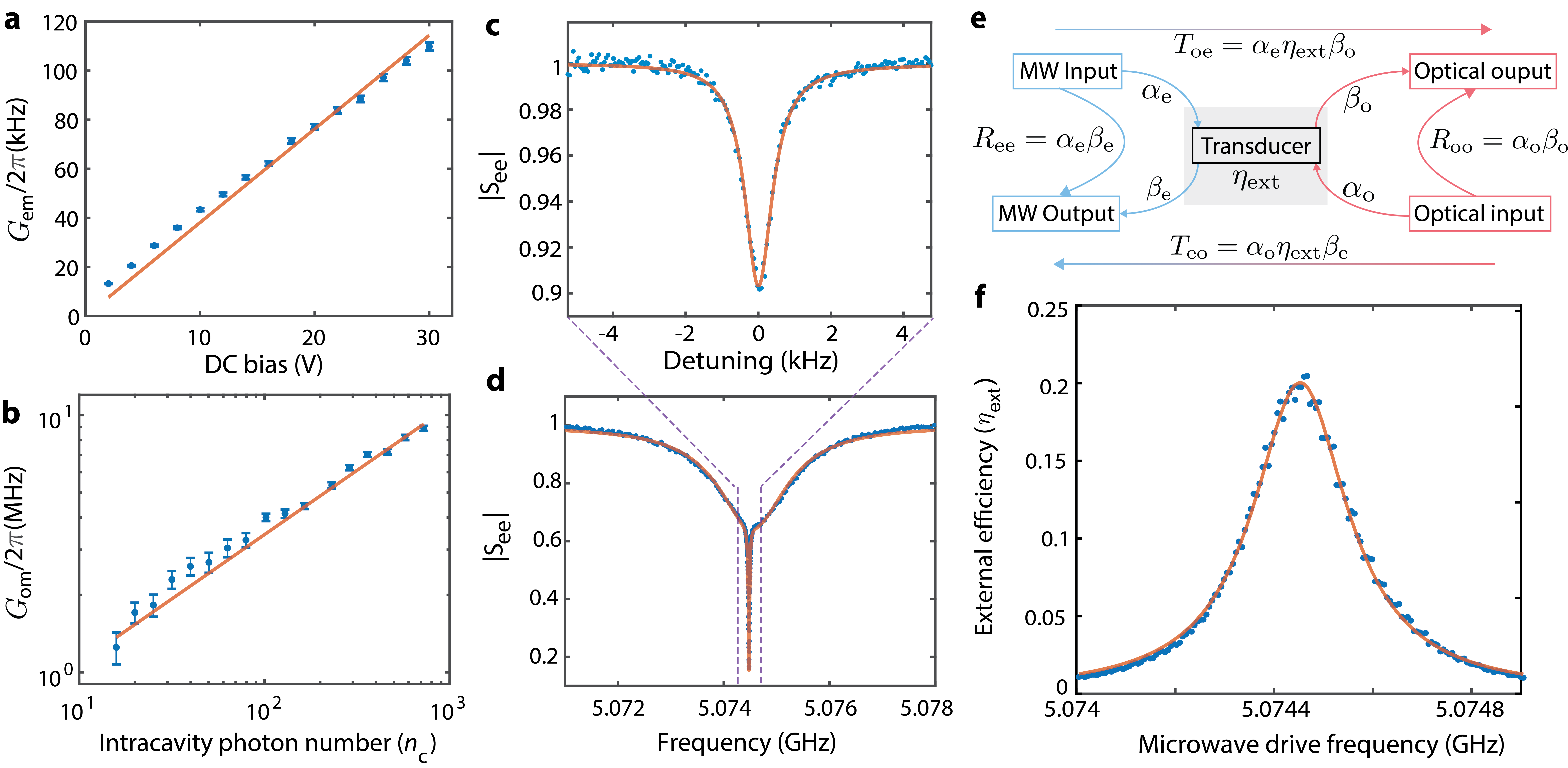}

\caption{\textbf{Transducer characterization} \textbf{a} Measured electromechanical coupling rate vs DC voltage bias. The coupling rate is determined at each voltage by fitting the external decay rate of the mechanical mode. \textbf{b} Measured optomechanical coupling rate versus intracavity photon number. The coupling rate is derived at each photon number from the difference in the mechanical linewidth under a red-detuned and a resonant pump with the same intracavity photon number. \textbf{c} Reflection spectrum of the mechanical resonance. The electromechanical external decay rate is minimized by choosing a large microwave-mechanics detuning of $31 \text{ MHz}$ at $V_{\text{DC}} = 40 \text{ V}$. The laser power is turned off, and the mechanical resonator is strongly driven with a microwave tone (see \cref{sec:TLS} for more details). \textbf{d} Reflection spectrum of the microwave resonator on-resonance with the mechanical mode. \textbf{e} Overview diagram of the 4 port measurements for calibrating the conversion efficiency. The microwave measurement setup loss (gain) is denoted by $\alpha_{\text{e}}$ ($\beta_{\text{e}}$). The optical measurement loss (gain) is given by $\alpha_{\text{o}}$ ($\beta_{\text{o}}$). \textbf{f} Measured microwave-to-optical transduction efficiency versus microwave drive frequency at $V_{\text{DC}} = 50 \text{ V}$ and $n_{\text{c}} = 232$. The solid line depicts the Lorentzian fit of the measured transduction spectrum. The error bars in \textbf{a} and \textbf{b} represent the 95\% confidence interval. }
\label{figure 2}
\end{figure*}

\section*{Transducer design}
\label{sec:theoretical_concept}

Our electro-optomechanical transducer features a gigahertz-frequency mechanical oscillator that is simultaneously coupled to a microwave resonator and an optical cavity (\cref{figure 1}a). We achieve the electromechanical coupling using electrostatic actuation, where the mechanical vibration of a DC-voltage-biased capacitor creates a time-dependent dipole moment oscillating at the mechanical frequency \cite{Bozkurt2023Sep}. The charged moving capacitor is connected to a microwave resonator that is resonant with the mechanical motion to facilitate efficient inter-conversion between the microwave photons and phonons. The mechanical oscillator also interacts with an optical resonator as part of a cavity optomechanical system \cite{Aspelmeyer2014Dec}, where a red-detuned optical pump converts phonons to optical photons. The system's response can be described via the Hamiltonian
\begin{equation}
\begin{split} 
\hat{H}/\hbar = & \omega_{\text{e}}\hat{a}_{\text{e}}^{\dagger}\hat{a}_{\text{e}} + \omega_{\text{m}}\hat{a}_{\text{m}}^{\dagger}\hat{a}_{\text{m}} + \omega_{\text{o}}\hat{a}_{\text{o}}^{\dagger}\hat{a}_{\text{o}} \\
+& G_{\text{em}}(\hat{a}_{\text{e}}\hat{a}_{\text{m}}^{\dagger} + \hat{a}_{\text{e}}^{\dagger}\hat{a}_{\text{m}}) + G_{\text{om}}(\hat{a}_{\text{o}}\hat{a}_{\text{m}}^{\dagger} + \hat{a}_{\text{o}}^{\dagger}\hat{a}_{\text{m}}). 
\label{eq:transduction efficiency}
\end{split}
\end{equation}
Here, $\omega_{{j}}/2\pi$ ($j =$ e,m,o) are the resonance frequencies of the microwave, mechanical and optical resonators, respectively. The electromechanical and optomechanical couplings ($G_{\text{em}} = g_{\text{em}}V_{\text{DC}}$, $G_{\text{om}} = g_{\text{om}}\sqrt{n_{\text{c}}}$) are enhanced from their `vacuum' values ($g_{\text{em}}$, $g_{\text{om}}$) via a biasing DC voltage ($V_{\text{DC}}$) and an intracavity photon number ($n_{\text{c}}$) created by the laser pump. 
%The symbols $\hat{a}_{{j}}$ and $\hat{a}_{{j}}^{\dagger}$ denote the corresponding annihilation and creation operators. 

We realize this system in an integrated device based on a suspended nanobeam of single-crystal silicon. The nanobeam is patterned with an air-hole phononic crystal structure containing an electromechanical (EMC) and an optomechanical crystal (OMC) cavity. The EMC section of the nanobeam supports a `breathing' mechanical mode and is situated between the electrodes of a parallel-plate capacitor that is connected to an on-chip microwave resonator. The OMC section of the nanobeam supports an optical mode with a strong overlap to the mechanical mode, providing optomechanical coupling through the photoelastic response in silicon \cite{Chan2012Aug}. To keep the optical components distant from the light-sensitive superconducting circuits, we separate the EMC and OMC sections by extending the mechanical mode across the nanobeam using a phononic waveguide (\cref{figure 1}b) \cite{Zhao2023Jun}.  We fabricate the transducer on a silicon-on-insulator (SOI) substrate along with a high-impedance flux-tunable microwave resonator made from niobium titanium nitride (NbTiN) (\cref{figure 1}c,d,e) \cite{Bretz-Sullivan2022Aug, Xu2019May}. 

 \begin{figure*}[htbp]
\centering
\includegraphics[width=\linewidth]{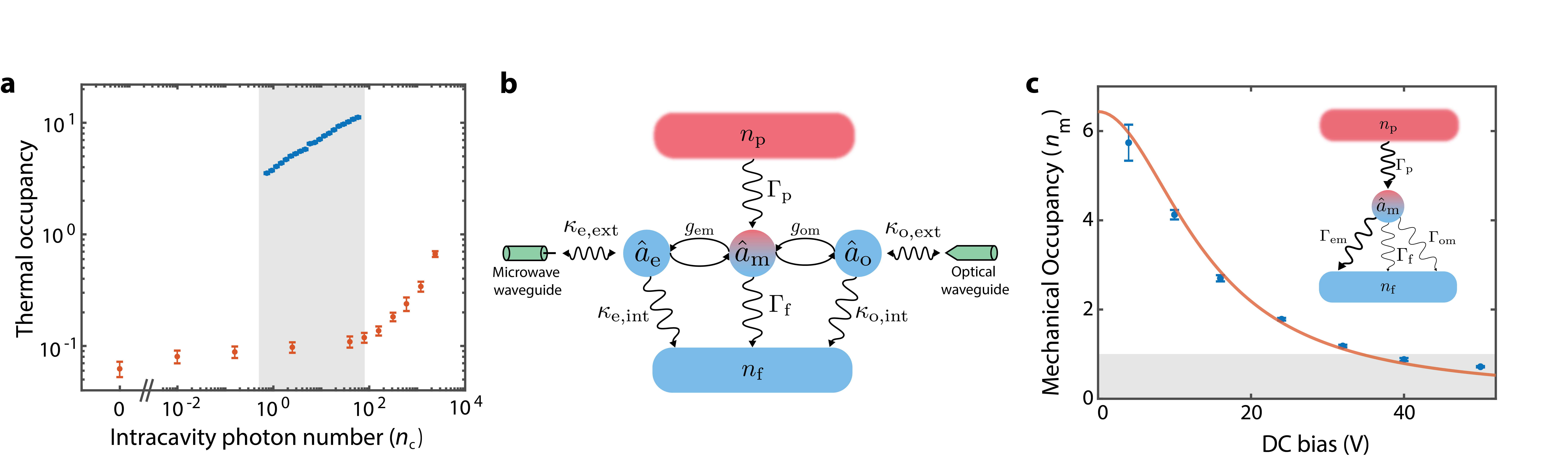}
\caption{\textbf{Mechanical quantum ground-state cooling.} \textbf{a} Thermometry measurements of the mechanical occupancy (blue circles) and microwave occupancy (red circles) at varying intracavity photon numbers. The shaded region idnicates the range of photon numbers in which the transducer is operated. The mechanical occupancy is measured with $\Gamma_{\text{em}}/2\pi = 2.1$ kHz, which is attained with $V_{\text{DC}} = 40$ V and a deliberate detuning of the microwave resonator from the mechanical resonance frequency by 4.4 MHz. \textbf{b} Diagram of the heating and damping model for the mechanical mode. 
 \textbf{c} Electromechanical cooling as a function of DC bias voltage. The red curve is the theoretical prediction (see \cref{sec:occupancy}). The measurements are performed with a continuous optical pump power corresponding to $n_{\text{c}} = 2.3$. Inset emphasizes the dominating cooling path via the electromechanical decay channel ($\Gamma_{\text{em}}>>\Gamma_{\text{f}},\Gamma_{\text{om}}$). Error bars in \textbf{a} and \textbf{c} represent the 95$\%$ confidence interval. In \textbf{a} and \textbf{c}, the mechanical occupancy is measured under the red-detuned optical pump. Due to the small $\Gamma_{\text{om}}$ compared to $\Gamma_{\text{i}}$, the optomechanical backaction cooling is negligible.}
\label{figure 3}
\end{figure*}

\section*{Transducer efficiency characterization}

We characterize the transducer by mounting the device to the mixing stage of a dilution refrigerator. We first probe the mechanical response using a pump laser red-detuned from the optical cavity. \Cref{figure 1}f shows the power spectral density of the optomechanical scattering, from which we identify two mechanical modes. To confirm the mechanical modes provide sufficient electromechanical coupling for transduction, we apply a DC bias voltage to the device and measure the coherent microwave-to-optical scattering. As shown in \cref{figure 1}g, we find two transduction peaks with the frequencies aligning with the mechanical modes in the optomechanical spectrum. To achieve single-mode operation, we tune the microwave resonator onto resonance with the mechanical mode at $\omega_\text{m}/2\pi = 5.074$ GHz, which has a better balance of electromechanical and optomechanical coupling rates. 

Our transducer operates in the Purcell regime where the microwave and optical resonator linewidths are greater than their respective coupling rates to mechanics \cite{SafaviNaeni2011Jan}. In this regime, the electrical and optical resonators serve as external radiation ports for the mechanical mode with the decay rates given by the parametrically enhanced coupling rates and the resonators' total decay rates: $\Gamma_{\text{em}} = 4G_{\text{em}}^2/\kappa_{\text{e}}$ and 
$\Gamma_{\text{om}} = 4G_{\text{om}}^2/\kappa_{\text{o}}$. The dependence of the electromechanical (optomechanical) decay rate on the DC voltage (laser power) provides a means of characterizing the interaction rate. Using this approach (see \cref{figure 2}a,b), we measure an electromechanical (optomechanical) interaction rate of $g_{\text{em}}/2\pi = 3.81 \text{ kHz/V}$ ($g_{\text{om}}/2\pi = 343 \text{ kHz}$), in reasonable agreement with device modeling \cite{Zhao2023Jun}. The extracted interaction rates and the mechanical internal linewidth can be used to calculate the bi-directional conversion efficiency. The transducer external efficiency, defined by the photon flux ratio between the microwave and optical waveguides on the chip, is given by (see \cref{sec:efficiency and bandwidth}) 
\begin{equation}
\eta_{\text{ext}} =  \eta_{\text{e}} \eta_{\text{o}} \frac{4\Gamma_{\text{em}}\Gamma_{\text{om}}}{(\Gamma_{\text{i}} +  \Gamma_{\text{em}} + \Gamma_{\text{om}})^2} .
\label{eq:transduction external efficiency}
\end{equation}
Here,  
$\eta_{\text{e}} = \kappa_{\text{e,ext}}/\kappa_{\text{e}}$ ($\eta_{\text{o}} = \kappa_{\text{o,ext}}/\kappa_{\text{o}}$) is the microwave (optical) resonator external coupling efficiency, and  $\Gamma_{\text{i}}$ denotes the intrinsic decay rates of the mechanical resonator. \Cref{figure 2}c shows the reflection spectrum of the mechanical resonator measured via the electrical port with the laser drive turned off. The fitted intrinsic linewidth, $\Gamma_{\text{i}}/2\pi = 892 \text{ Hz}$ ($Q_\mathrm{m} =5.7 \times 10^6$), provides an estimate for the mechanical decay rate in our system, and its exceptionally small value is a key feature of our platform. However, this value cannot be used to directly calculate the transduction efficiency due to the dependence of the mechanical linewidth on the laser drive in our system \cite{Meenehan2014Jul} (see \cref{sec:TLS}). 

 As an alternative, we directly characterize the efficiency by measuring the reflectance ($R_\text{ee}$, $R_{\text{oo}}$) and transmittance ($T_\text{oe}$, $T_{\text{eo}}$) between the transducer's microwave and optical ports \cite{Andrews2014Apr}. The calculated quantity from this experiment $\eta_{\text{ext}} = \sqrt{T_\text{oe}T_\text{eo}/(R_\text{oo}R_\text{ee})}$ provides the transducer external efficiency with no need for calibrating the gains and losses in the measurement paths (\cref{figure 2}e) (see \cref{sec:4-port}). Using this method, we obtain a maximum external efficiency of $\eta_{\text{ext,max}} = 20.0\% \pm 0.9\%$ (see \cref{figure 2}f). In our system, this value is limited by the maximum DC voltage allowed by the measurement apparatus ($V_{\text{DC}} = 50$ V). We also calculate the efficiency by directly measuring the mechanical linewidth with the laser power on and using \cref{eq:transduction external efficiency}. The resulting value $\eta_{\text{ext,pred}} = 22.0 \pm 1.5 \%$, agrees with the result from four-port characterization, which verifies our model of the device efficiency.  As detailed in the next section, achieving the quantum-enabled regime in our device requires a different set of conditions, resulting in smaller efficiencies. Regardless, we point out that the maximum efficiency is quite large compared to other integrated mechanics-mediated transducers \cite{Jiang2023Oct,Weaver2024Feb}, despite our system's lower electromechanical interaction rates, which signifies the benefits of the low mechanical loss.

\begin{figure*}[htbp]
\centering
\includegraphics[width=0.95\linewidth]{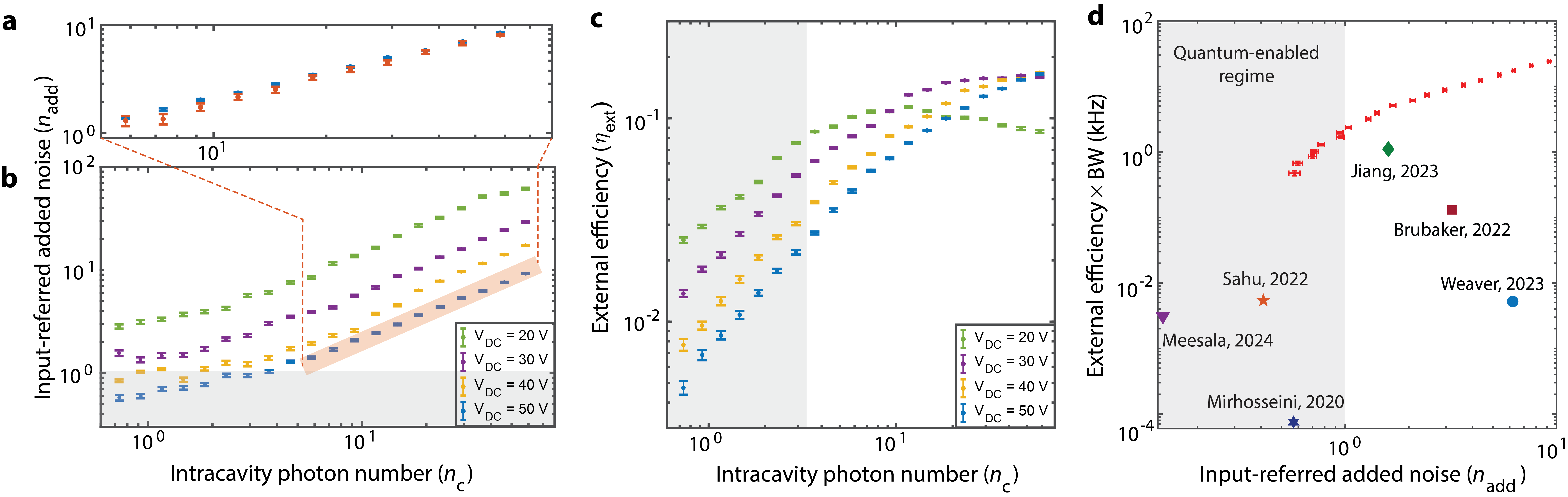}
\caption{\textbf{Quantum-enabled microwave-to-optical transduction.} \textbf{a} Electrically (blue points) and optically (orange points) measured input-referred added noise of the transducer with varying optical intracavity photon number. \textbf{b} Electrically measured input-referred added noise of the transducer with varying optical intracavity photon numbers. \textbf{c} Corresponding transduction efficiencies for the same photon numbers and bias voltages as in \textbf{b} from four-port measurements. The shaded region represents the quantum-enabled regime at 50 V biasing DC voltage. \textbf{d} Comparison of the transducer performance with the literature. Red dots are the measured performance of our transducer. References: Meesala, 2024 \cite{Meesala2024Feb}; Sahu, 2022 \cite{Sahu2022Mar}; Mirhosseini, 2020 \cite{Mirhosseini2020Dec}; Jiang, 2023 \cite{Jiang2023Oct}; Brubaker, 2022 \cite{brubaker_optomechanical_2022}, Weaver, 2023 \cite{Weaver2024Feb}. Error bars in all panels represent the 95$\%$ confidence interval. Shaded regions in \textbf{b} and \textbf{c} denote the quantum-enabled regime, where $n_{\text{add}} < 1$.} 
\label{figure 4}
\end{figure*}

 \section*{Quantum ground-state cooling} 

 Thermal occupation in the microwave and mechanical resonators is the primary source of noise in an electro-optomechanical transducer \cite{Meenehan2014Jul, Meenehan2015Oct, Mirhosseini2020Dec, Jiang2023Oct, Meesala2024Feb, Sahu2022Mar, Weaver2024Feb}. To characterize the effective mode temperature of these components in our device, we perform thermometry measurements via the electrical port (see \cref{sec:Electrical thermometry with gain calibration}). \Cref{figure 3}a shows the measurement results. We observe that the microwave thermal population remains low at $n_{\text{mw}}\lesssim 0.1$ and approximately independent of the laser power for $n_{\text{c}} \lesssim 200$ (equivalent to $3.3 \text{ }\mu\text{W}$ at the optical waveguide).  We attribute the microwave resonator's robustness against laser drive to the fast quasiparticle recombination rate of NbTiN \cite{Lobo2005Jul} and the optical waveguide design that we have used to minimize direct illumination of the superconductor (\cref{figure 1}c). In contrast, we observe a much larger mechanical occupation with a pronounced dependence on laser power. This observed heating of the mechanical mode is consistent with previous measurements in a variety of integrated optomechanical devices  \cite{Meenehan2014Jul, Meenehan2015Oct, Stockill2019Oct, Ren2020Jul, Ramp2019Aug} and is the major roadblock to achieving low-noise transduction.

The mechanical heating can be understood via a phenomenological model (see \cref{figure 3}b), where a laser-power-dependent hot bath (at the occupation number $n_{\text{p}}$) is coupled to the mechanical mode (at the rate $\Gamma_{\text{p}}$) \cite{Meenehan2014Jul}. The heating from this hot bath competes with the cooling rate of the mechanical oscillator by the fridge's cold bath ($\Gamma_{\text{f}}$) and the electrical and optical output ports ($\Gamma_{\text{em}} +  \Gamma_{\text{om}}$). At equilibrium,  the mechanical occupation can be found as \footnote{In this expression, we neglect the contribution to mechanical heating from the nonzero microwave occupation to simplify the presentation. The full model used for all our data analysis captures this effect and is described in supplementary section ID.}
\begin{equation}
n_{\text{m}} \approx \frac{ \Gamma_{\text{p}} n_{\text{p}} } { \Gamma_{\text{i}} + \Gamma_{\text{em}} +  \Gamma_{\text{om}} }.
\label{eq:mechanical occupation}
\end{equation}
Here, $\Gamma_{\text{i}} =\Gamma_{\text{f}}+\Gamma_{\text{p}} $ is the intrinsic mechanical decay rate. The dependence of mechanical occupation on the electromechanical and optomechanical decay rates provides a means of mitigating thermal noise via these interactions. In our system, the electromechanical decay rate can greatly exceed all the other rates, acting as the dominant source of cooling. \Cref{figure 3}c shows the results of an experiment where the mechanical occupation is measured for a fixed laser power (corresponding to $n_{\text{c}} = 2.3$) as we gradually increase the electromechanical decay rate by increasing $V_{\text{DC}}$. It can be seen that the mechanical mode, which is initially quite hot, enters the quantum ground state ($n_{\text{m}}<1$) for sufficiently large $V_{\text{DC}}$. 

We observe that the ground-state electromechanical cooling is only possible because of the absence of any observable heating from the electrostatic field. This observation is supported by the good agreement between our measurements and theoretical predictions, as shown in \cref{figure 3}c. This contrasts with processes that utilize radio-frequency pumps, which can create significant heating \cite{Kalaee2019Apr}. Moreover, the effectiveness of electromechanical cooling in mitigating heating is largely due to the inherently low heating rate in our system. This rate, determined by the product $\Gamma_{\text{p}}n_{\text{p}}$, depends on both the optical absorption coefficient and the mechanical loss tangent, properties that are strongly material-dependent \cite{Jiang2019Jul, Stockill2019Oct, Forsch2020Jan, Joe2023Oct, Ramp2019Aug}. Previous studies have identified silicon as having low optical absorption as well as exceptionally low acoustic loss \cite{Ren2020Jul, MacCabe2020Nov}. The material-independent nature of electrostatic coupling in our approach thus allows us to leverage these advantageous properties, together with radiative cooling, to achieve low-noise operation as we elaborate on in the next section.

\section*{Quantum-enabled operation} 

Although ground-state cooling is a necessary condition for quantum-enabled operation, it is not sufficient for it. To find the added noise for operation in up- or down-conversion (i.e., microwave-to-optics or optics-to-microwave), the thermal occupations from all components need to be referred to the corresponding input port. In our system, the much larger electromechanical decay rate $\Gamma_{\text{em}} \gg  \Gamma_{\text{om}}$ creates asymmetric implications for the added noise in these two operation modes, where we can only reach the quantum-enabled regime in upconversion. This asymmetric performance can be understood by the fact that electromechanical cooling channels thermal noise primarily to the microwave port, with the optical output port remaining largely noise-free.

We utilize two independent methods for measuring the input-referred added noise $n_{\text{add}}$  in microwave-to-optics conversion. In the first approach, the noise in the optical output port ($n_{\text{out}}$) is measured directly in a heterodyne measurement setup and then referred to the input using the external efficiency found from the four-port measurements described above, $ n_{\text{add}} = {n_{\text{out}}}/{\eta_{\text{ext}}}$ (see \cref{sec:optical noise}). In the second approach, we electrically measure the mechanical thermal occupancy and convert it to $n_{\text{add}}$ via the relation $n_{\text{add}} = n_{\text{m}} { \Gamma_{\text{tot}} }/( {\eta_{\text{e}}\Gamma_{\text{em}} })$ (see \cref{sec:Added noise}).
\Cref{figure 4}a compares the results from these two measurement methods. Although we find good agreement between the two methods, we use the electrical measurements in the low-noise regime because of its much higher signal-to-noise ratio, which results from the much larger readout rate in the electrical port. \Cref{figure 4}b plots the electrical measurement results for various values of laser power and voltage bias. As evident, for $V_{\text{DC}}=$ 50 V, the input noise drops below $n_{\text{add}} = 1$ for an extended range of laser powers. 

\Cref{figure 4}c shows the transduction efficiencies from four-port measurements for the voltage and photon number values used in \cref{figure 4}b. The shaded region in the figure depicts the parameter space where the added noise remains below one for $V_{\text{DC}} = $ 50 V. We observe that increasing $\Gamma_{\text{em}}$ at a fixed optical pump power results in reducing the noise as well as the efficiency, where the increasing imbalance between the electromechanical and optomechanical decay rates causes the latter. However, the transduction efficiency increases with increasing $\Gamma_{\text{em}}$  for any fixed value of added noise due to increased cooperativities. Constrained to operate in the quantum-enabled regime, our transducer achieves a maximum external efficiency of $\eta_{\text{ext}} = 2.2\%$ at $n_{\text{add}} = 0.94$. At this operation setting, the transducer bandwidth, given by the total mechanical linewidth, is measured as $\text{B} = \Gamma_{\text{tot}}/2\pi = 88.9 \pm 2.1$ kHz.

\section*{Discussion} 
Our demonstration of quantum-enabled performance in continuous operation contrasts all previous demonstrations, which have reached this regime via pulsed operation. Conceptually, continuous operation has the advantage of utilizing the entire efficiency-bandwidth product, whereas pulsed operation is penalized by the duty cycle. Nevertheless, this conceptual distinction does not provide a complete picture, as different platforms provide vastly different bandwidths. To provide a quantitative measure of our results and provide proper context with respect to previous work, we use the transducer throughput, defined as the product of efficiency $\eta_{\text{ext}}$, bandwidth B, and duty cycle D, where $\text{D} = 1$ for continuous operation.  In \cref{figure 4}d, we plot this product versus the input referred added noise from our measurements and previous experiments. In the quantum-enabled regime, our transducer achieves $\eta_{\text{ext}} \times \text{B} \times \text{D} = 1.9 \text{ kHz}$ with $n_{\text{add}} = 0.94$. We also observe a trade-off between noise and throughput in our data. At our lowest noise setting, we obtain $n_{\text{add}} = 0.58$ and $\eta_{\text{ext}} \times \text{B} \times \text{D} = 470 \text{ Hz}$, which is about two orders of magnitude larger than reported values in \cite{Sahu2022Mar}. Although lowering the optical pump power is expected to reduce noise further, accurately characterizing the reduced device efficiency in this regime was prohibitively long with heterodyne measurements in our experimental setup.  Further investigation of our platform in this regime with single-photon detectors will be the subject of future work.

\section*{Conclusions and outlook}

In summary, we demonstrate quantum-enabled microwave-to-optical frequency conversion on an integrated silicon platform. Utilizing electromechanical ground-state cooling, our transducer operates continuously and achieves a throughput that surpasses previous work significantly. The enhanced photon emission rates resulting from this improvement hold great promise for future demonstrations of optical remote entanglement with superconducting qubits. Moreover, the simplicity of fabricating our devices, which rely solely on standard silicon-on-insulator substrates and metal deposition, is promising for scalable adoption, potentially via foundry services. Importantly, our demonstrated bandwidth adequately matches the coherence of standard planar qubits, and continuous operation simplifies integration with a qubit source, which can be placed on a separate chip and connected by low-loss coaxial cables \cite{10.1038/s41928-023-00925-z}. A technical consideration for these future experiments would be isolating the qubit source from the thermal noise emanating from the electrical port of the transducer, which can be achieved either by low-loss bulk circulators or by engineering directional qubit-waveguide interfaces \cite{10.1038/s41567-022-01869-5,Chaitali2023}. Lastly, we expect our system to improve in noise performance by about an order of magnitude using optomechanical cavities with better thermal contact \cite{Ren2020Jul,10.1364/optica.492143}. With reduced noise, the demonstrated kilohertz-level throughputs would enable quantum networking operations with superconducting qubits at rates comparable to optical qubit platforms \cite{Bernien:2013kj,10.1126/science.aan5959}. 

\section*{Acknowledgments}
We acknowledge S. Meesala, O. Painter, A. Safavi-Naeini, A. Bozkurt, and M Kalaee for helpful discussions. This work was supported by the ARO/LPS Cross Quantum Technology Systems program (grant W911NF-18-1-0103), the US Department of Energy Office of Science National Quantum Information Science Research Centers (Q-NEXT, award DE-AC02-06CH11357), and National Science Foundation (awards 2137645 and 2238058).  W.D.C gratefully acknowledges support from the National Science Foundation Graduate Research Fellowship.

% \bibliography{references}

\clearpage

\newpage

\appendix
\onecolumngrid
\section{Theoretical model of electro-optomechanical transducers}
\label{sec:theory}
\subsection{Hamiltonian}

Our electro-optomechanical transducer is comprised of a mechanical resonator that is simultaneously coupled to a microwave and optical resonator. The bare Hamiltonian is given by
\begin{equation}
\frac{\hat{H}_{\text{o}}}{\hbar} = \omega_{\text{e}}\hat{a}_{\text{e}}^\dagger\hat{a}_{\text{e}} + \omega_{\text{m}}\hat{a}_{\text{m}}^\dagger\hat{a}_{\text{m}} +  \omega_{\text{o}}\hat{a}_{\text{o}}^\dagger\hat{a}_{\text{o}},
\label{eq:bare hamiltonian}
\end{equation}
where $\omega_{j}$ , with $j = \text{e,m,o}$, are the microwave, mechanical, and optical resonator frequencies, respectively, and $\hat{a}_{j}^\dagger$ ($\hat{a}_{j}$) are the corresponding raising (lowering) operators. Note that $\hat{a}_{\text{o}}$ represents fluctuations about the coherent optical pump field.   

To simplify the following analysis, we work with the optical mode in a rotating frame at the pump laser frequency $\omega_{\text{L}}$. The bare Hamiltonian becomes 

\begin{equation}
\frac{\hat{H}_{\text{o}}}{\hbar} = \omega_{\text{e}}\hat{a}_{\text{e}}^\dagger\hat{a}_{\text{e}} + \omega_{\text{m}}\hat{a}_{\text{m}}^\dagger\hat{a}_{\text{m}} +  \Delta_{\text{o}}\hat{a}_{\text{o}}^\dagger\hat{a}_{\text{o}},
\label{eq:bare hamiltonian rotating}
\end{equation}
where $\Delta_{\text{o}} = \omega_{\text{o}} - \omega_{\text{L}}$ is the pump laser detuning. 

The electromechanical interaction term is given by \cite{Bozkurt2023Sep}
\begin{equation}
\begin{split}
\frac{\hat{H}_{\text{em}}}{\hbar} & = \frac{\partial{C_{\text{m}}}}{\partial{x}}x_{\text{zpf}}V_{\text{zpf}}V_{\text{DC}}(\hat{a}_{\text{m}}^\dagger\hat{a}_{\text{e}} + \hat{a}_{\text{m}}\hat{a}_{\text{e}}^\dagger) \\
 & = G_{\text{em}}(\hat{a}_{\text{m}}^\dagger\hat{a}_{\text{e}} + \hat{a}_{\text{m}}\hat{a}_{\text{e}}^\dagger),
\end{split}
\label{eq:em interaction hamiltonian}
\end{equation}
where $C_{\text{m}} $ is the capacitance of the moving capacitor, $x_{\text{zpf}}$ is the zero point motion of the mechanical resonator, $V_{\text{zpf}}$ is the zero point voltage of the microwave resonator, and $V_{\text{DC}}$ is the DC bias applied to the capacitor. $G_{\text{em}} = \frac{\partial{C_{\text{m}}}}{\partial{x}}x_{\text{zpf}}V_{\text{zpf}}V_{\text{DC}}$ is the electromechanical coupling rate.     

The optomechanical interaction term is given by \cite{Aspelmeyer2014Dec}
\begin{equation}
\begin{split}
\frac{\hat{H}_{\text{om}}}{\hbar} = g_{\text{o}}\sqrt{n_{\text{c}}}(\hat{a}_{\text{o}}^\dagger + \hat{a}_{\text{o}})(\hat{a}_{\text{m}}^\dagger + \hat{a}_{\text{m}}) 
 \approx G_{\text{om}}(\hat{a}_{\text{o}}^\dagger\hat{a}_{\text{m}} + \hat{a}_{\text{o}}\hat{a}_{\text{m}}^\dagger),
\end{split}
\label{eq:om interaction hamiltonian}
\end{equation}
where $G_{\text{om}} = g_{\text{o}}\sqrt{n_{\text{c}}}$ is the optomechanical coupling rate. The interaction takes on a beam splitter form since we operate with the pump laser red-detuned from the optical cavity ($\Delta_{\text{o}} = \omega_{\text{m}}$) in the sideband resolved regime ($\omega_{\text{m}} >> \kappa_{\text{o}}$). 

Therefore, the full Hamiltonian is given by 
\begin{equation}
\begin{split}
\frac{\hat{H}}{\hbar} =  \omega_{\text{e}}\hat{a}_{\text{e}}^\dagger\hat{a}_{\text{e}} + \omega_{\text{m}}\hat{a}_{\text{m}}^\dagger\hat{a}_{\text{m}} +  \Delta_{\text{o}}\hat{a}_{\text{o}}^\dagger\hat{a}_{\text{o}} + G_{\text{em}}(\hat{a}_{\text{m}}^\dagger\hat{a}_{\text{e}} + \hat{a}_{\text{m}}\hat{a}_{\text{e}}^\dagger) + G_{\text{om}}(\hat{a}_{\text{o}}^\dagger\hat{a}_{\text{m}} + \hat{a}_{\text{o}}\hat{a}_{\text{m}}^\dagger).
\end{split}
\label{eq:full hamiltonian}
\end{equation}

\subsection{Heisenberg-Langevin equations of motion}
By introducing noise operators into the Heisenberg equations of motion, we can write down a system of first-order differential equations that fully describe the response of the electro-optomechanical transducer in the presence of thermal baths. The noise correlations are given by
\begin{equation}
\begin{split}
\langle \hat{a}_{j,\text{ext}}^\dagger(t) \hat{a}_{j,\text{ext}}(t') \rangle & = \langle \hat{a}_{j,\text{ext}}(t) \hat{a}_{j,\text{ext}}^\dagger(t') \rangle - \delta(t - t')  =  n_{j,\text{ext}}\delta(t - t'), \\
\langle \hat{a}_{j,\text{int}}^\dagger(t) \hat{a}_{j,\text{int}}(t') \rangle & = \langle \hat{a}_{j,\text{int}}(t) \hat{a}_{j,\text{int}}^\dagger(t') \rangle - \delta(t - t')  = n_{j,\text{int}}\delta(t - t'), \\
\langle \hat{a}_{\text{f}}^\dagger(t) \hat{a}_{\text{f}}(t') \rangle & = \langle \hat{a}_{\text{f}}(t) \hat{a}_{\text{f}}^\dagger(t') \rangle - \delta(t - t')  = n_{\text{f}}\delta(t - t'), \\
\langle \hat{a}_{\text{p}}^\dagger(t) \hat{a}_{\text{p}}(t') \rangle & = \langle \hat{a}_{\text{p}}(t) \hat{a}_{\text{p}}^\dagger(t') \rangle - \delta(t - t')  = n_{\text{p}}\delta(t - t'),
\end{split}
\label{eq:noise correlations}
\end{equation}
where $\hat{a}_{j,\text{ext}}$, with $j = $ e,o are the electrical and optical waveguide baths with damping rate $\kappa_{j,\text{ext}}$ and thermal occupancy $n_{j,\text{ext}}$ ; $\hat{a}_{j,\text{int}}$ are the intrinsic resonator baths with damping rate $\kappa_{j,\text{int}}$ and thermal occupancy $n_{j,\text{int}}$ ; $\hat{a}_{\text{f}}$ is the intrinsic fridge mechanics bath with damping rate $\Gamma_{\text{f}}$ and thermal occupancy $n_{\text{f}}$ ; $\hat{a}_{\text{p}}$ is a hot mechanics bath due to optical absorption heating with damping rate $\Gamma_{\text{p}}$ and thermal occupancy $n_{\text{p}}$.\\

The corresponding Heisenberg-Langevin equations of motion are given by
\begin{equation}
\begin{split}
&\dot{\hat{a}}_{\text{o}} = -(i\Delta_{\text{o}} + \kappa_{\text{o}}/2)\hat{a}_{\text{o}} - iG_{\text{om}}\hat{a}_{\text{m}}
+ \sqrt{\kappa_{\text{o,ext}}}\hat{a}_{\text{o,ext}}+ \sqrt{\kappa_{\text{o,int}}}\hat{a}_{\text{o,int}},\\
&\dot{\hat{a}}_{\text{e}} = -(i\omega_{\text{e}} + \kappa_{\text{e}}/2)\hat{a}_{\text{e}} - iG_{\text{em}}\hat{a}_{\text{m}} 
 + \sqrt{\kappa_{\text{e,ext}}}\hat{a}_{\text{e,ext}} + \sqrt{\kappa_{\text{e,int}}}\hat{a}_{\text{e,int}}, \\
 &\dot{\hat{a}}_{\text{m}} = -(i\omega_{\text{m}} + \Gamma_{\text{i}}/2)\hat{a}_{\text{m}} - iG_{\text{om}}\hat{a}_{\text{o}} - iG_{\text{em}}\hat{a}_{\text{e}} 
+ \sqrt{\Gamma_{\text{f}}}\hat{a}_{\text{f}} + \sqrt{\Gamma_{\text{p}}}\hat{a}_{\text{p}},
\end{split}
\label{eq: equations of motion}
\end{equation}
where $\kappa_{j}= \kappa_{j,\text{ext}} + \kappa_{j,\text{int}}$ is the resonator total linewidth, and $\Gamma_{\text{i}} = \Gamma_{\text{f}} + \Gamma_{\text{p}}$ is the mechanical intrinsic linewidth.  

From input-output theory, the output microwave and optical fields are given by $\hat{a}_{j,\text{out}} = \sqrt{\kappa_{j,\text{ext}}}\hat{a}_{j} - \hat{a}_{j,\text{ext}}$.\\

This system of first-order differential equations can be described using a state-space model \cite{Andrews2014Apr, Arnold2020Sep}
\begin{equation}
\begin{split}
\dot{a}(t) = Aa(t) + Ba_{\text{in}}(t) , \\
a_{\text{out}}(t) = Ca(t) + Da_{\text{in}}(t),
\end{split}
\label{eq:input ouput}
\end{equation}
where $a_{\text{out}} = [\hat{a}_{\text{e,out}}, \hat{a}_{\text{o,out}}, \hat{a}_{\text{e,out}}^\dagger, \hat{a}_{\text{o,out}}^\dagger]^\intercal$ contains the output fields, 
$a_{\text{in}} = [\hat{a}_{\text{e,ext}}, \hat{a}_{\text{e,int}}, \hat{a}_{\text{o,ext}}, \hat{a}_{\text{o,int}},\hat{a}_{\text{f}},\hat{a}_{\text{p}}\\,\hat{a}_{\text{e,ext}}^\dagger,\hat{a}_{\text{e,int}}^\dagger, \hat{a}_{\text{o,ext}}^\dagger, \hat{a}_{\text{o,int}}^\dagger, \hat{a}_{\text{f}}^\dagger,\hat{a}_{\text{p}}^\dagger]^\intercal$ contains the input fields, and $a = [\hat{a}_{\text{e}}, \hat{a}_{\text{o}}, \hat{a}_{\text{m}}, \hat{a}_{\text{e}}^\dagger, \hat{a}_{\text{o}}^\dagger, \hat{a}_{\text{m}}^\dagger]^\intercal$ contains the resonator fields. The matrices are given by
\begin{equation}
A = \begin{pmatrix}
-(i\omega_{\text{e}} + \kappa_{\text{e}}/2) & 0 & -iG_{\text{em}} & 0 & 0 & 0\\
0 & -(i\Delta_{\text{o}} + \kappa_{\text{o}}/2) & -iG_{\text{om}} & 0 & 0 & 0\\
-iG_{\text{em}} & -iG_{\text{om}} & -(i\omega_{\text{m}} + \Gamma_{\text{i}}/2) & 0 & 0 & 0\\
0 & 0 & 0 & -(-i\omega_{\text{e}} + \kappa_{\text{e}}/2) & 0 & iG_{\text{em}}\\
0 & 0 & 0 & 0 & -(-i\Delta_{\text{o}} + \kappa_{\text{o}}/2) & iG_{\text{om}}\\
0 & 0 & 0 & iG_{\text{em}} & iG_{\text{om}} & -(-i\omega_{\text{m}} + \Gamma_{\text{i}}/2)\\
\end{pmatrix},
\end{equation}

\begin{equation}
B = \begin{pmatrix}
\sqrt{\kappa_{\text{e,ext}}} & \sqrt{\kappa_{\text{e,int}}} & 0 & 0 & 0 & 0 & 0 & 0 & 0 & 0 & 0 & 0\\
0 & 0 & \sqrt{\kappa_{\text{o,ext}}} & \sqrt{\kappa_{\text{o,int}}} & 0 & 0 & 0 & 0 & 0 & 0 & 0 & 0\\
0 & 0 & 0 & 0 & \sqrt{\Gamma_{\text{f}}} & \sqrt{\Gamma_{\text{p}}} & 0 & 0 & 0 & 0 & 0 & 0\\
0 & 0 & 0 & 0 & 0 & 0 & \sqrt{\kappa_{\text{e,ext}}} & \sqrt{\kappa_{\text{e,int}}} & 0 & 0 & 0 & 0\\
0 & 0 & 0 & 0 & 0 & 0 & 0 & 0 & \sqrt{\kappa_{\text{o,ext}}} & \sqrt{\kappa_{\text{o,int}}} & 0 & 0\\
0 & 0 & 0 & 0 & 0 & 0 & 0 & 0 & 0 & 0 & \sqrt{\Gamma_{\text{f}}} & \sqrt{\Gamma_{\text{p}}}\\
\end{pmatrix},
\end{equation}

\begin{equation}
C = \begin{pmatrix}
\sqrt{\kappa_{\text{e,ext}}} & 0 & 0 & 0 & 0 & 0\\
0 & \sqrt{\kappa_{\text{o,ext}}} & 0 & 0 & 0 & 0\\
0 & 0 & 0 & \sqrt{\kappa_{\text{e,ext}}} & 0 & 0\\
0 & 0 & 0 & 0 & \sqrt{\kappa_{\text{o,ext}}} & 0\\
\end{pmatrix},
\end{equation}

\begin{equation}
D = \begin{pmatrix}
-1 & 0 & 0 & 0 & 0 & 0 & 0 & 0 & 0 & 0 & 0 & 0\\
0 & 0 & -1 & 0 & 0 & 0 & 0 & 0 & 0 & 0 & 0 & 0\\
0 & 0 & 0 & 0 & 0 & 0 & -1 & 0 & 0 & 0 & 0 & 0\\
0 & 0 & 0 & 0 & 0 & 0 & 0 & 0 & -1 & 0 & 0 & 0\\
\end{pmatrix}.
\end{equation}

The transfer function of the electro-optomechanical transducer relates the input to output fields in the Fourier domain: 
$a_{\text{out}}(\omega) = \xi(\omega)a_{\text{in}}(\omega)$ where $\xi(\omega) = C(-i\omega I-A)^{-1}B+D$ and $I$ is the identity matrix. 

\subsection{Efficiency and bandwidth}
\label{sec:efficiency and bandwidth}
\label{efficiency and bandwidth theory}
The microwave-to-optical transmission coefficient $s_{\text{oe}}(\omega)$ can be calculated from the transfer function matrix element that maps input field $\hat{a}_{\text{e,ext}}$ to output field $\hat{a}_{\text{o,out}}$:
\begin{equation}
    s_{\text{oe}}(\omega) = -\frac{\sqrt{\eta_{\text{e}}}\sqrt{\eta_{\text{o}}}\sqrt{\kappa_{\text{e}}}\sqrt{\kappa_{\text{o}}}G_{\text{em}}\chi_{\text{e}}(\omega)G_{\text{om}}\chi_{\text{o}}(\omega)}{G_{\text{em}}^2\chi_{\text{e}}(\omega) + G_{\text{om}}^2\chi_{\text{o}}(\omega) + \chi_{\text{m}}^{-1}(\omega)},     
\end{equation}
where $\eta_{j} = \kappa_{j,\text{ext}}/\kappa_{j}$ for $j = \text{e,o}$, and the susceptibilities are given by $\chi_{\text{o}}^{-1} = i(\Delta_{\text{o}} - \omega) + \kappa_{\text{o}}/2$, $\chi_{\text{e}}^{-1} = i(\omega_{\text{e}} - \omega) + \kappa_{\text{e}}/2$, and $\chi_{\text{m}}^{-1} = i(\omega_{\text{m}} - \omega) + \Gamma_{\text{i}}/2$. \newline

Since the device operates in the weak-coupling regime ($G_{\text{em}} <<  \kappa_{\text{e}}$ and $G_{\text{om}} <<  \kappa_{\text{o}}$), we can express the transmission coefficient around the mechanical mode in terms of the resonator-mediated damping rates $\Gamma_{\text{em}} = \frac{G_{\text{em}}^2\kappa_{\text{e}}}{(\omega_{\text{e}}-\omega_{\text{m}})^2 + (\kappa_{\text{e}}/2)^2}$ and $\Gamma_{\text{om}} = \frac{G_{\text{om}}^2\kappa_{\text{o}}}{(\Delta_{\text{o}}-\omega_{\text{m}})^2 + (\kappa_{\text{o}}/2)^2}$:  

\begin{equation}
    s_{\text{oe}}(\omega) =  -\frac{\sqrt{\eta_{\text{e}}}\sqrt{\eta_{\text{o}}}\sqrt{\Gamma_{\text{em}}}\sqrt{\Gamma_{\text{om}}}}{i(\omega_{\text{m}} - \omega) + (\Gamma_{\text{em}} + \Gamma_{\text{om}} + \Gamma_{\text{i}})/2}.
\end{equation}
Note that $|s_{\text{oe}}(\omega)| = |s_{\text{eo}}(\omega)|$ since the beam-splitter interaction is symmetric with respect to microwave-to-optical and optical-to-microwave state conversion. 

Thus, the external microwave-optics conversion efficiency is given by
\begin{equation}
    \eta_{\text{ext}}(\omega) = |s_{\text{oe}}(\omega_{\text{m}})|^2 = |s_{\text{eo}}(\omega_{\text{m}})|^2 = \eta_{\text{e}}\eta_{\text{o}}\frac{\Gamma_{\text{em}}\Gamma_{\text{om}}}{(\omega_{\text{m}} - \omega)^2 + \left((\Gamma_{\text{em}} + \Gamma_{\text{om}} + \Gamma_{\text{i}})/2\right)^2 },
\label{eq:external efficiency definition}
\end{equation}
and the internal transduction efficiency is $\eta_{\text{int}}(\omega) = \eta_{\text{ext}}(\omega)/(\eta_{\text{e}}\eta_{\text{o}})$. \newline

In the on-resonance condition ($\omega_{\text{e}} = \Delta_{\text{o}} = \omega = \omega_{\text{m}}$), the peak external efficiency is as reported in the main text:\\
\begin{equation}
\eta_{\text{ext}} = \eta_{\text{e}}\eta_{\text{o}}\frac{4\Gamma_{\text{em}}\Gamma_{\text{om}}}{(\Gamma_{\text{i}} +  \Gamma_{\text{em}} + \Gamma_{\text{om}})^2} = \eta_{\text{e}}\eta_{\text{o}}\frac{4\Gamma_{\text{em}}\Gamma_{\text{om}}}{\Gamma_{\text{tot}}^2}, 
\label{eq:transduction efficiency}
\end{equation}
where $\Gamma_{\text{em}} = 4G_{\text{em}}^2/\kappa_{\text{e}}$,  $\Gamma_{\text{om}} = 4G_{\text{om}}^2/\kappa_{\text{o}}$, and $\Gamma_{\text{tot}} = \Gamma_{\text{i}} + \Gamma_{\text{em}} + \Gamma_{\text{om}}$.

From \cref{eq:external efficiency definition}, it is clear that the transducer bandwidth is given by the total mechanical linewidth: $\text{B} = \Gamma_{\text{tot}}/2\pi = (\Gamma_{\text{em}} + \Gamma_{\text{om}} + \Gamma_{\text{i}})/2\pi$. 

\subsection{Microwave and mechanical resonator thermal occupancy}
\label{sec:occupancy}
The microwave and mechanical resonator thermal occupancy can be derived from the Heisenberg-Langevin equations of motion \cref{eq: equations of motion}. In the Fourier domain, the equations of motion become

\begin{equation}
\begin{split}
&\hat{a}_{\text{o}}(\omega) = (-iG_{\text{om}}\hat{a}_{\text{m}}(\omega)
+ \sqrt{\kappa_{\text{o,ext}}}\hat{a}_{\text{o,ext}}(\omega)+ \sqrt{\kappa_{\text{o,int}}}\hat{a}_{\text{o,int}}(\omega))\chi_{\text{o}}(\omega),\\
&\hat{a}_{\text{e}}(\omega) =  (-iG_{\text{em}}\hat{a}_{\text{m}} 
 + \sqrt{\kappa_{\text{e,ext}}}\hat{a}_{\text{e,ext}}(\omega) + \sqrt{\kappa_{\text{e,int}}}\hat{a}_{\text{e,int}}(\omega))\chi_{\text{e}}(\omega), \\
 &\hat{a}_{\text{m}}(\omega) = (-iG_{\text{om}}\hat{a}_{\text{o}}(\omega) - iG_{\text{em}}\hat{a}_{\text{e}}(\omega) 
+ \sqrt{\Gamma_{\text{f}}}\hat{a}_{\text{f}} + \sqrt{\Gamma_{\text{p}}}\hat{a}_{\text{p}})\chi_{\text{m}}(\omega).
\end{split}
\label{eq: fourier equations of motion}
\end{equation} 

Since the mechanical resonator operates in the weak coupling regime ($G_{\text{em}} << \kappa_{\text{e}}$), $\hat{a}_{\text{e}}(\omega)$ can be written as 

\begin{equation}
\hat{a}_{\text{e}}(\omega) =  (\sqrt{\kappa_{\text{e,ext}}}\hat{a}_{\text{e,ext}}(\omega) + \sqrt{\kappa_{\text{e,int}}}\hat{a}_{\text{e,int}}(\omega))\chi_{\text{e}}(\omega). \\
\end{equation} 

For our device, the thermal baths that contribute to the resonator thermal occupancies are $n_{\text{e,int}}$, $n_{\text{f}}$, and $n_{\text{p}}$ since $n_{\text{e,ext}} \approx n_{\text{o,ext}} \approx n_{\text{o,int}} \approx 0$ (see comments in \cref{sec:Electrical thermometry with gain calibration}). Therefore, the microwave resonator thermal occupancy is given by \cite{Marquardt2007Aug}

\begin{equation}
\begin{split}
n_{\text{mw}} &= \frac{1}{2\pi}\int_{-\infty}^{+\infty}d\omega\int_{-\infty}^{+\infty}d\omega' \langle \hat{a}^\dagger_{\text{e}}(\omega') \hat{a}_{\text{e}}(\omega)\rangle = \frac{\kappa_{\text{e,int}}n_{\text{e,int}}}{\kappa_{\text{e}}}.
\end{split}
\label{eq: mw occupancy}
\end{equation}

Similarly, the mechanical resonator thermal occupancy is given by 

\begin{equation}
\begin{split}
n_{\text{m}} &= \frac{1}{2\pi}\int_{-\infty}^{+\infty}d\omega\int_{-\infty}^{+\infty}d\omega' \langle \hat{a}^\dagger_{\text{m}}(\omega') \hat{a}_{\text{m}}(\omega)\rangle =  \frac{\Gamma_{\text{em}}n_{\text{mw}} + \Gamma_{\text{f}}n_{\text{f}} + \Gamma_{\text{p}}n_{\text{p}}}{\Gamma_{\text{tot}}}.
\end{split}
\label{eq: mechanics occupancy}
\end{equation}

\subsection{Added noise}
\label{sec:Added noise}
The transducer's added noise in microwave-to-optical and optical-to-microwave frequency conversion due to thermal baths can be calculated from the transfer function matrix elements that map thermal bath inputs to output field $\hat{a}_{\text{o,out}}$ and $\hat{a}_{\text{e,out}}$, respectively.

Using the transfer matrix $\xi(\omega)$, we find the output optical field $\hat{a}_{\text{o,out}}(\omega)$ in the presence of all thermal baths to be

\begin{equation}
\begin{split}
    \hat{a}_{\text{o,out}}(&\omega) = -\frac{\sqrt{\eta_{\text{e}}}\sqrt{\eta_{\text{o}}}\sqrt{\Gamma_{\text{em}}}\sqrt{\Gamma_{\text{om}}}}{i(\omega_{\text{m}} - \omega) + \Gamma_{\text{tot}}/2}\left(\hat{a}_{\text{e,ext}}(\omega) + \sqrt{\frac{\kappa_{\text{e,int}}}{\kappa_{\text{e,ext}}}}\hat{a}_{\text{e,int}}(\omega)\right) -\frac{i\sqrt{\eta_{\text{o}}}\sqrt{\Gamma_{\text{om}}}}{i(\omega_{\text{m}} - \omega) + \Gamma_{\text{tot}}/2}\left(\sqrt{\Gamma_{\text{f}}}\hat{a}_{\text{f}}(\omega) + \sqrt{\Gamma_{\text{p}}}\hat{a}_{\text{p}}(\omega)\right)\\& 
    + \left(\frac{i(\omega_{\text{m}} - \omega) + (\Gamma_{\text{em}} + \Gamma_{\text{i}})/2}{i(\omega_{\text{m}} - \omega) + \Gamma_{\text{tot}}/2} \kappa_{\text{o,ext}}\chi_{\text{o}}(\omega)  - 1\right)\hat{a}_{\text{o,ext}}(\omega) + \frac{i(\omega_{\text{m}} - \omega) + (\Gamma_{\text{em}} + \Gamma_{\text{i}})/2}{i(\omega_{\text{m}} - \omega) + \Gamma_{\text{tot}}/2} \sqrt{\kappa_{\text{o,ext}}}\sqrt{\kappa_{\text{o,int}}}\chi_{\text{o}}(\omega)\hat{a}_{\text{o,int}}(\omega).
\end{split}
\end{equation}

Similarly, we can find the output microwave field $\hat{a}_{\text{e,out}}(\omega)$ to be 

\begin{equation}
\label{electrical output field}
\begin{split} 
    \hat{a}_{\text{e,out}}(&\omega) = -\frac{\sqrt{\eta_{\text{e}}}\sqrt{\eta_{\text{o}}}\sqrt{\Gamma_{\text{em}}}\sqrt{\Gamma_{\text{om}}}}{i(\omega_{\text{m}} - \omega) + \Gamma_{\text{tot}}/2}\left(\hat{a}_{\text{o,ext}}(\omega) + \sqrt{\frac{\kappa_{\text{o,int}}}{\kappa_{\text{o,ext}}}}\hat{a}_{\text{o,int}}(\omega)\right) -\frac{i\sqrt{\eta_{\text{e}}}\sqrt{\Gamma_{\text{em}}}}{i(\omega_{\text{m}} - \omega) + \Gamma_{\text{tot}}/2}\left(\sqrt{\Gamma_{\text{f}}}\hat{a}_{\text{f}}(\omega) + \sqrt{\Gamma_{\text{p}}}\hat{a}_{\text{p}}(\omega)\right)\\& 
    + \left(\frac{i(\omega_{\text{m}} - \omega) + (\Gamma_{\text{om}} + \Gamma_{\text{i}})/2}{i(\omega_{\text{m}} - \omega) + \Gamma_{\text{tot}}/2} \kappa_{\text{e,ext}}\chi_{\text{e}}(\omega)  - 1\right)\hat{a}_{\text{e,ext}}(\omega) + \frac{i(\omega_{\text{m}} - \omega) + (\Gamma_{\text{om}} + \Gamma_{\text{i}})/2}{i(\omega_{\text{m}} - \omega) + \Gamma_{\text{tot}}/2} \sqrt{\kappa_{\text{e,ext}}}\sqrt{\kappa_{\text{e,int}}}\chi_{\text{e}}(\omega)\hat{a}_{\text{e,int}}(\omega).
\end{split}
\end{equation}

We analyze the electrical output noise in our discussion of electrical thermometry (see \cref{sec:Electrical thermometry with gain calibration}). For the rest of this section, we focus on the noise at the optical output for microwave-to-optical conversion. The optical noise spectral density $n_{\text{o,out}}(\omega)$ is related to the output optical field by 

\begin{equation}
 n_{\text{o,out}}(\omega) = \int_{-\infty}^{\infty} d\tau e^{i\omega\tau} \langle \hat{a}_{\text{o,out}}^\dagger(0) \hat{a}_{\text{o,out}}(\tau) \rangle = \int_{-\infty}^{\infty} d\omega' \langle (\hat{a}_{\text{o,out}}(\omega'))^\dagger \hat{a}_{\text{o,out}}(\omega) \rangle .
\end{equation}

 For our device, the thermal baths that contribute to the added noise are $n_{\text{e,int}}$, $n_{\text{f}}$, and $n_{\text{p}}$ since $n_{\text{e,ext}} \approx n_{\text{o,ext}} \approx n_{\text{o,int}} \approx 0$ (see comments in \cref{sec:Electrical thermometry with gain calibration}). Therefore, the thermal noise at the optical output is

\begin{equation}
\label{optical spectral density}
\begin{split}
    n_{\text{o,out}}(\omega) = 
    \frac{\eta_{\text{e}}\eta_{\text{o}}\Gamma_{\text{em}}\Gamma_{\text{om}}}{(\omega_{\text{m}} - \omega)^2 + (\Gamma_{\text{tot}}/2)^2}\frac{\kappa_{\text{e,int}}}{\kappa_{\text{e,ext}}}n_{\text{e,int}} &+\frac{\eta_{\text{o}}\Gamma_{\text{om}}}{(\omega_{\text{m}} - \omega)^2 + (\Gamma_{\text{tot}}/2)^2}\Gamma_{\text{f}}n_{\text{f}}\\& 
    +\frac{\eta_{\text{o}}\Gamma_{\text{om}}}{(\omega_{\text{m}} - \omega)^2 + (\Gamma_{\text{tot}}/2)^2}\Gamma_{\text{p}}n_{\text{p}} .
\end{split}
\end{equation}

The up-conversion input-referred added noise is
    
\begin{equation}
\begin{split}
    n_{\text{add}} = n_{\text{o,out}}(\omega)/\eta_{\text{ext}}(\omega) = 
    \frac{\kappa_{\text{e,int}}}{\kappa_{\text{e,ext}}}n_{\text{e,int}} +\frac{\Gamma_{\text{f}}n_{\text{f}} + \Gamma_{\text{p}}n_{\text{p}}}{\eta_{\text{e}}\Gamma_{\text{em}}}.
\end{split}
\end{equation}

We can express the input-referred added noise in terms of the mechanical mode occupancy:

\begin{equation}
\begin{split}
    n_{\text{add}} = 
    \frac{\Gamma_{\text{em}}n_{\text{mw}} + \Gamma_{\text{f}}n_{\text{f}} + \Gamma_{\text{p}}n_{\text{p}}}{\eta_{\text{e}}\Gamma_{\text{em}}} = n_{\text{m}}\frac{\Gamma_{\text{tot}}}{\eta_{\text{e}}\Gamma_{\text{em}}}.
\end{split}
\label{eq:n_add from n_m}
\end{equation}
where $n_{\text{mw}} = \frac{\kappa_{\text{e,int}}}{\kappa_{\text{e}}}n_{\text{e,int}}$ is the microwave resonator thermal occupancy (see \cref{eq: mw occupancy}), and $n_{\text{m}} = \frac{\Gamma_{\text{em}}n_{\text{mw}} + \Gamma_{\text{f}}n_{\text{f}} + \Gamma_{\text{p}}n_{\text{p}}}{\Gamma_{\text{tot}}}$ is the mechanical mode thermal occupancy (see \cref{eq: mechanics occupancy}).

\newpage

\section{Fabrication}
We fabricate the transducers starting with a high resistivity ($> 3 \text{ k}\Omega\cdot \text{cm}$) 220-nm silicon-on-insulator substrate sputtered with a 12-nm thin film of niobium titanium nitride (NbTiN) with sheet inductance of $\sim 36$ pH/sq. NbTiN is used as the superconductor for microwave circuits due to its large kinetic inductance and fast quasi-particle recombination rate \cite{Bretz-Sullivan2022Aug, Lobo2005Jul}. First, we selectively remove the NbTiN film from the optical and mechanical structures and define the microwave superconducting circuit by electron-beam lithography (EBL), followed by a pseudo-Bosch dry etch with SF$_6$/Ar chemistry. This step is followed by a second aligned EBL and  SF$_6$/C$_4$F$_8$ dry etch to define the silicon nanobeams, phonon shields, release holes, and optical waveguides. To create space for lensed-fiber end-fire coupling, we etch a $\sim120$ $\mu$m deep trench area at the end of the optical waveguide using photolithography and SF$_6$ chemical etching. The devices are finally released with hydrofluoric (HF) acid and wire-bonded to a printed circuit board for electrical contact.

\section{Measurement setup}
\label{sec:Measurement setup}

\begin{figure}[htbp]
\centering
\includegraphics[width=0.9\columnwidth]{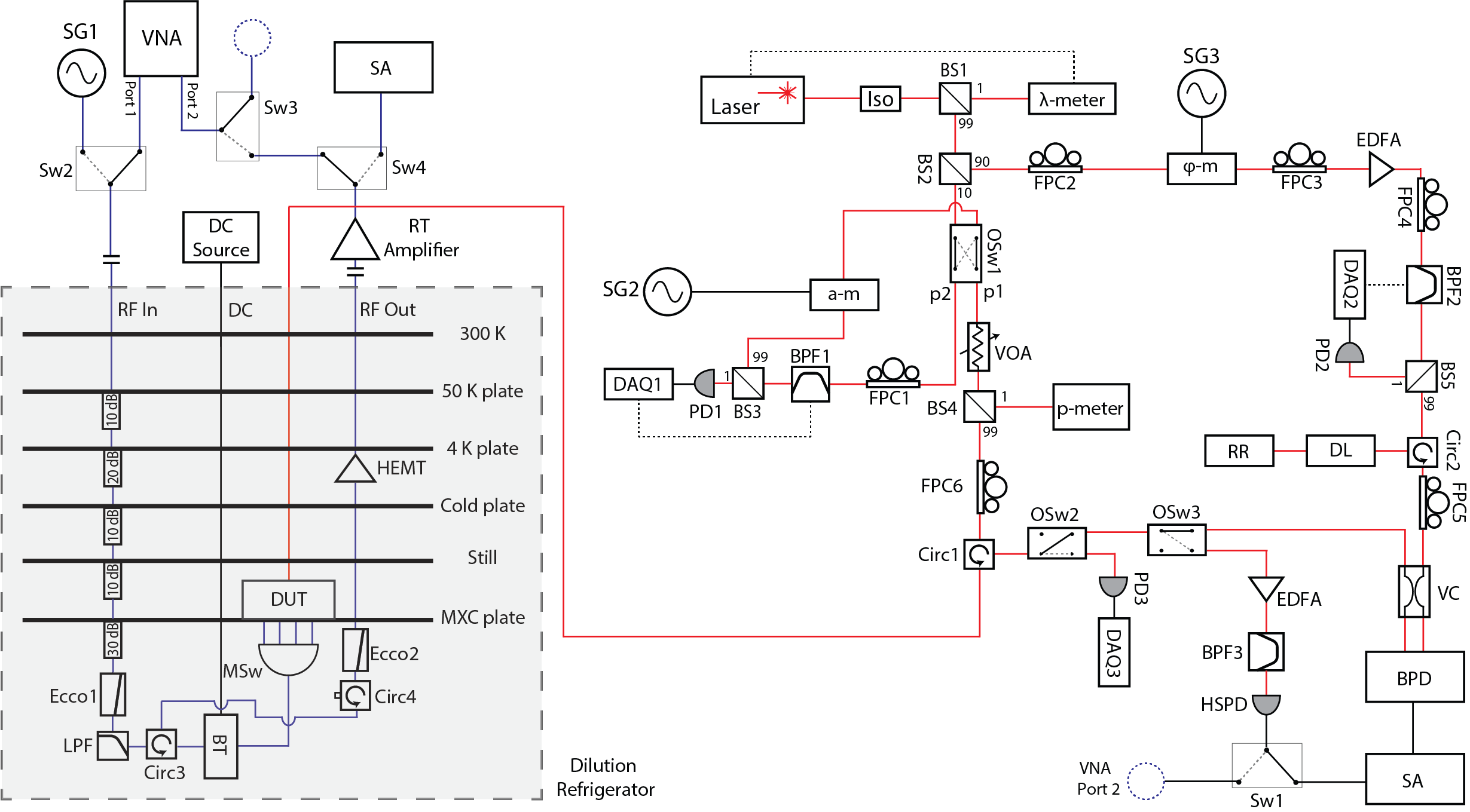}
\caption{\textbf{Measurement setup.} Acronyms: Isolator (Iso), beam splitter (BS), wavelength meter ($\lambda$-meter), optical switch (OSw), fiber polarization controller (FPC), band-pass filter (BPF), photodetector (PD), data-acquisition module (DAQ), amplitude modulator (a-m), signal generator (SG), variable optical attenuator (VOA), power meter (p-meter), circulator (Circ), device under test (DUT), phase modulator ($\phi$-m), Erbium-doped fiber amplifier (EDFA), delay line (DL), retro-reflector (RR), variable coupler (VC), high-speed photodetector (HSPD), balanced photodetector (BPD), spectrum analyzer (SA), switch (Sw), vector network analyzer (VNA), eccosorb filter (Ecco), low-pass filter (LPF), bias-tee (BT), microwave switch (MSw), high-electron-mobility transistor (HEMT), room-temperature amplifier (RT amplifier).
}
\label{measurement setup}
\end{figure}

\begin{figure}[htbp]
\centering
\includegraphics[width=0.5\columnwidth]{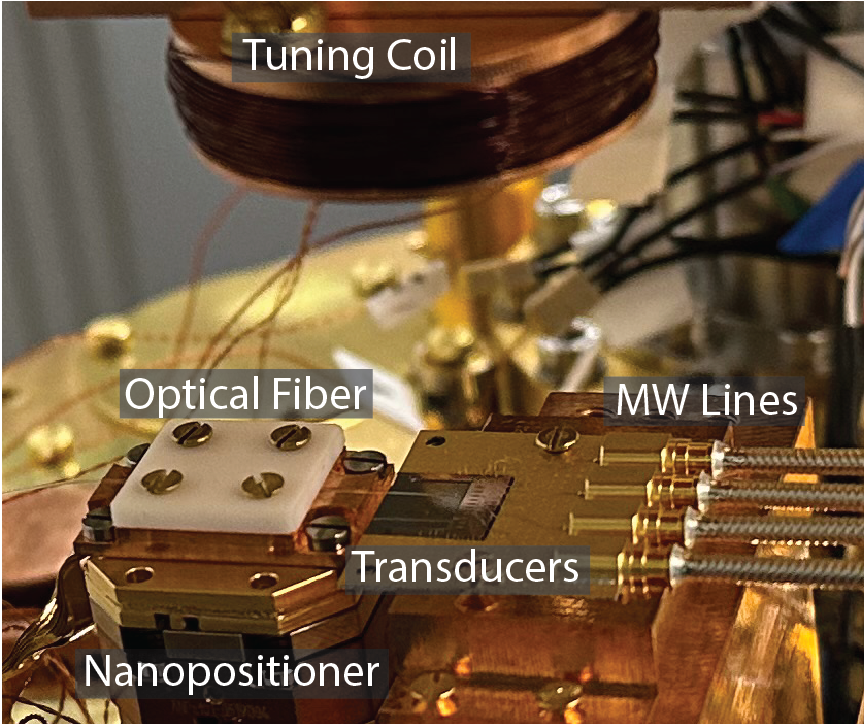}
\caption{\textbf{Fridge setup}. A tuning coil is placed above the device chip to generate magnetic flux that tunes the on-chip microwave resonators. The device chip sits on a printed circuit board (PCB), with the on-chip coplanar waveguides wire-bonded to the PCB traces. The microwave lines are connected to the PCB via MMPX connectors.  
}
\label{figure:fridge setup}
\end{figure}

\Cref{measurement setup} illustrates the optics and microwave setup used for all measurements. A tunable laser provides the optical pump for the transducer and is frequency-stabilized using a wavelength meter. For the transduction measurements, the optical pump beam is filtered by a 50-MHz tunable bandpass filter (BPF) to suppress laser phase noise at the $\sim$5-GHz sidebands of the laser frequency \cite{Safavi-Naeini2013}. An amplitude modulator (a-m) is used to generate a probe tone at the optical resonance frequency as needed. The optical pump power is controlled by a variable optical attenuator (VOA) and is sent to the device under test (DUT) at the mixing stage of the dilution refrigerator after an optical circulator (Circ1). A lensed optical fiber mounted on a set of cryogenic x-y-z nanopositioners is used to couple to the optical waveguides on the DUT (\cref{figure:fridge setup}). 

The optical reflection from the DUT contains the reflected pump, the transducer's output signal, and the added noise photons. We characterize the optical reflection from the DUT using both homodyne and heterodyne detection setups. In the homodyne setup, the reflected signal is amplified by an Erbium-doped fiber amplifier (EDFA) and then detected by a high-speed photodetector (HSPD). Before the HSPD, a tunable bandpass filter (BPF3) is used to reject the excessive spontaneous emission from the EDFA. The beating notes of the pump reflection and the transduction output generate photocurrents at the mechanical frequencies. We send the photocurrents to a spectrum analyzer (SA) and a vector network analyzer (VNA) to characterize the optomechanical spectrum and the microwave-to-optical transduction spectrum, respectively. The sensitivity of the homodyne detection, however, is too low for accurate characterizations of the transducer's efficiency and noise performance as the optical pump power decreases.

For the measurements of the external efficiency and added noise, we switch to the heterodyne setup where the transduction signal can be amplified by a higher-intensity optical local oscillator (LO) independent of the optical pump power. In this setup, the LO is split from the laser power using a 90:10 beam splitter (BS2), with 90\% of the power directed to the LO arm. The LO beats with the optical reflection from the DUT via a variable coupler (VC), which is subsequently converted to photocurrents by a balanced photodetector (BPD). To generate the beating notes within the bandwidth of the BPD (400 MHz) for the transduction signal, the LO is blue-shifted by a frequency 50 MHz lower than the mechanical resonances. We achieve the frequency shift by a chain of phase modulator ($\phi$-m), EDFA and another 50-MHz tunable bandpass filter (BPF2). The optical length of LO is matched to the reflection signal from the DUT by a round-trip delay line (DL) to minimize the decoherence of the beating notes. We use the SA to analyze the transduction and the added noise contained in the photocurrents from the BPD.

The microwave input signal is provided by either a VNA or a signal generator (SG). To minimize the room-temperature microwave noise, the input signal is attenuated by a total of $\sim$ 80 dB, with the attenuators thermalized to different stages of the dilution refrigerator. The attenuated microwave input passes through an Eccosorb filter (Ecco1) and a low-pass filter (LPF) to filter out high-frequency noise. A microwave circulator (Circ3) directs the input signal to the DUT and the reflection to the output line. The biasing DC voltage is applied to the DUT along with the microwave input via a bias-tee (BT). We use a microwave switch (MSw) to toggle the connections to different devices on the chip. An additional circulator (Circ4) and Eccosorb filter (Ecco2) are used at the output line to prevent the thermal noise of higher temperature stages from reaching the DUT. A high-electron-mobility transistor (HEMT) thermalized to the 4K stage, along with a low noise amplifier at room temperature (RT Amplifier), amplifies the microwave output of the DUT. The output signal after this amplification chain is detected by the SA or the VNA based on the type of measurements performed.

\newpage

\section{Transducer characterization}

\subsection{Device parameters}

\begin{table}[h!]
\begin{tabular}{|c|c|c|}
\hline
 \bf{Symbol} & \bf{Description} & \bf{Value} \\ 
  \hline
   $\omega_{\text{e}}/2\pi$ & Microwave resonator frequency & $5.0745 \text{ GHz}$ \\ 
  \hline
   $\kappa_{\text{e,ext}}/2\pi$ & Microwave resonator extrinsic linewidth & $1.33 \text{ MHz}$*\\
   \hline
   $\kappa_{\text{e,int}}/2\pi$ & Microwave resonator intrinsic linewidth & $330 \text{ kHz}$\\ 
   \hline
   $\omega_{\text{o}}/2\pi$ & Optical resonator frequency & $192.9263 \text{ THz}$ \\ 
   \hline
   $\kappa_{\text{o,ext}}/2\pi$ & Optical resonator extrinsic linewidth & $1.35 \text{ GHz}$ \\
   \hline
   $\kappa_{\text{o,int}}/2\pi$ & Optical resonator intrinsic linewidth & $404 \text{ MHz}$\\ 
   \hline
   $\omega_{\text{m}}/2\pi$ & Mechanical resonator frequency & $5.0745 \text{ GHz}$ \\ 
   \hline
   $\Gamma_{\text{i}}/2\pi$ & Mechanical mode saturated intrinsic linewidth & $892 \text{ Hz}$ \\
   \hline
   $g_{\text{em}}/2\pi$ & Electromechanical coupling rate & $3.81 \text{ kHz/V}$\\ 
   \hline
   $g_{\text{om}}/2\pi$ & Optomechanical single-photon coupling rate & $343 \text{ kHz}$\\ 
 \hline
 $\eta_{\text{f}}$ & Optical fiber-waveguide coupling efficiency & $35 \%$\\ 
 \hline
\end{tabular} \\
\hfill \break
* See \cref{Microwave resonator linewidth} for comments.
\caption{Device parameters}
\label{device parameters}
\end{table}    

\subsection{Microwave resonator linewidth}
\label{Microwave resonator linewidth}
The microwave resonator's external coupling rate, $\kappa_{\text{e,ext}}$, exhibits a weak, nonmonotonic dependence on the applied flux, with noticeable hysteresis. We attribute this to impedance mismatches caused by parasitic coupling between the microwave waveguide and neighboring resonators, which undergo phase slips at strong flux biases due to their large kinetic inductance \cite{Masluk2012Sep}. As a result of this behavior and the significant flux biasing applied before taking the dataset at $V_{\text{DC}} = 50 \text{ V}$, the measured value of  $\kappa_{\text{e,ext}}/2\pi = 1.33 \text{ MHz}$ at this voltage differs slightly from the value in all other experiments ($\kappa_{\text{e,ext}}/2\pi = 1.5 \text{ MHz}$). 

\subsection{Microwave resonator under optical illumination}
\label{append:microwave robustness}
\begin{figure}[h!]
\centering
\includegraphics[width =\columnwidth]{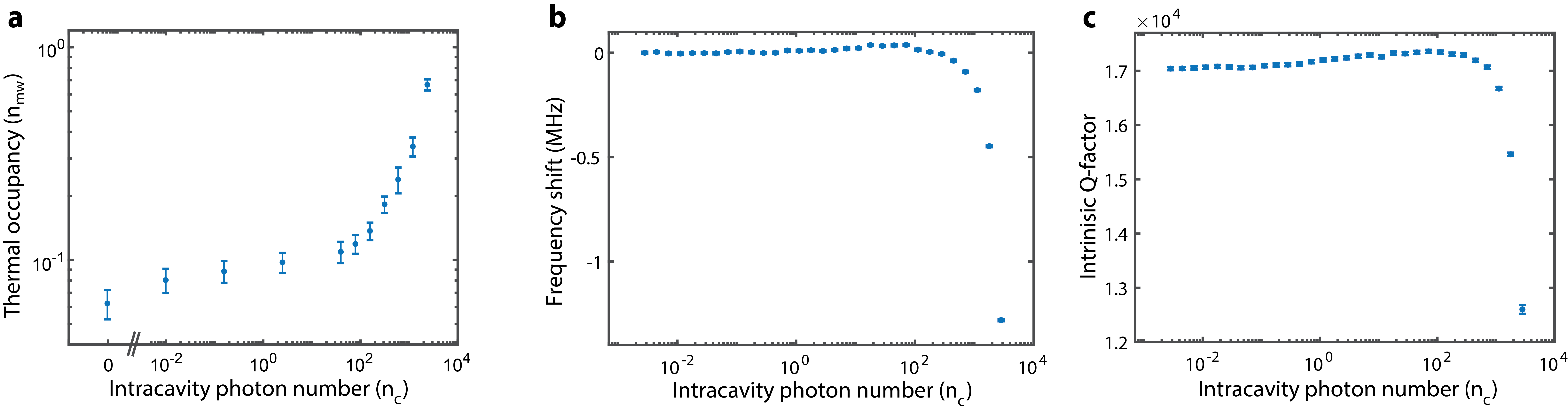}
\caption{\textbf{Microwave resonator under optical illumination.} a) Microwave resonator thermal occupancy vs intracavity photon number. The laser frequency is detuned by the mechanical frequency to the red sideband of the optical resonance for $n_{\text{c}} < 400$. The laser is tuned on-resonance with the optical resonator for $n_{\text{c}} > 400$ due to limited cooling power at the mixing plate. b) Microwave resonator frequency shift vs intracavity photon number. c) Microwave resonator intrinsic quality factor vs intracavity photon number. All data in (b,c) is collected with the laser on-resonance with the optical cavity. Error bars denote 95\% confidence intervals.   
}
\label{figure mw resonator optical}
\end{figure}

Optical laser illumination of superconductors can break Cooper pairs via the absorption of high-energy optical photons. The resulting non-equilibrium quasiparticle population is believed to cause the increased microwave thermal occupancy, degradation of the resonator Q-factor, and frequency shift observed in previous transduction works \cite{Mirhosseini2020Dec, Weaver2024Feb, Xu2024Jan}. These effects on the microwave resonator are undesired since they directly lead to increased noise and lower transduction efficiency. Our transducer's NbTiN nanowire microwave resonator is shown to remain robust under high intracavity optical pump photon numbers, which we attribute to the superconductor's fast quasiparticle recombination rate \cite{Lobo2005Jul} and an optical waveguide design that minimizes direct illumination of the superconductor (see Figure 2c of main text). \Cref{figure mw resonator optical} shows the microwave thermal occupancy, frequency shift, and intrinsic Q-factor of the microwave resonator as a function of the intracavity optical pump photon number. For $n_{\text{c}} < 200$, we measure the microwave resonator thermal occupancy to be $n_{\text{mw}} \approx 0.1$ . The corresponding microwave temperature ($\sim100$ millikelvin) is reasonably close to the fridge temperature ($\sim30$ millikelvin), which is likely caused by reduced infrared shielding in our measurement setup. Significant heating of the microwave resonator, linewidth broadening, and frequency shift only occur for $n_{\text{c}} > 200$, a range that is two orders of magnitude greater than the pump photon numbers used for operating in the quantum-enabled regime.

\subsection{Mechanical resonator linewidth}
\label{sec:TLS}
\begin{figure}[htbp]
\centering
\includegraphics[width=0.35\columnwidth]{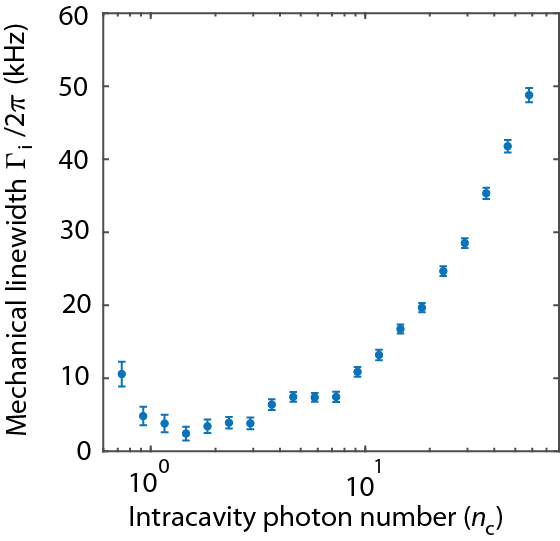}
\caption{\textbf{Mechanical linewidth versus optical pump power.} The intrinsic linewidth, calculated by subtracting the electromechanical and optomechanical decay rates from the measured total linewidth ($\Gamma_{\text{i}} = \Gamma_{\text{tot}} - \Gamma_{\text{om}} - \Gamma_{\text{em}}$). The microwave drive has been set to zero in this measurement, and the mechanical linewidth is measured from the thermal emission spectrum collected via the electrical port. The DC bias voltage is set at $V_{\text{DC}} = 30$ V. Error bars denote 95\% confidence intervals.
}
\label{fig:mech linewidth}
\end{figure}

The laser power strongly modifies the mechanical linewidth in our device (see \cref{fig:mech linewidth}). At low optical powers (corresponding to $n_{\text{c}} \lesssim 1$), the mechanical linewidth narrows as the optical pump power is increased.  In contrast, at higher optical power (corresponding to $n_{\text{c}} \gtrsim 5-10$), we observe a broadening of the mechanical linewidth concomitant with increased mechanical occupation. This behavior is consistent with previous observations in silicon phononic crystal resonators \cite{Meenehan2015}. The linewidth narrowing at low power has been attributed to the saturation of frequency jitter caused by two-level system (TLS) defects \cite{Wallucks2020Jul,MacCabe2020Nov,Bozkurt2023Sep}. The broadening at higher laser powers is a signature of optical absorption-induced mechanical decay, a process that has been attributed to 3-phonon mixing \cite{Meenehan2015,MacCabe2020Nov}. 

In the low optical power regime, where the mechanical linewidth is dominated by interaction with the TLS bath, the mechanical linewidth can also change as a function of the microwave drive \cite{Bozkurt2023Sep}. This behavior can be utilized to estimate the decay rate by saturating the frequency jitter via a microwave drive tone.  Figure 2c of the main text shows the results of such a measurement, where we populate the mechanical mode with $9.3 \times 10^5$ phonons in the absence of any optical pump. From a Lorentzian fit to the data, we measure a linewidth of $\Gamma_{\text{i}}/2\pi = 892 \text{ Hz}$. This measurement provides a conservative estimate for the decay rate ($\Gamma_{\text{f}}$) to the cold fridge bath ($n_{\text{f}})$.  For transduction efficiency and noise measurements presented in the main text, we operate in the regime where the frequency jitter has been nearly completely saturated with the laser power, and consequently, the device response becomes independent of the microwave drives (see  \cref{sec:Limitations} for details).

\subsection{Measurement of transduction efficiency}
\subsubsection{Four-port measurements}
\label{sec:4-port}
\begin{figure}[htbp]
\centering
\includegraphics[width=\columnwidth]{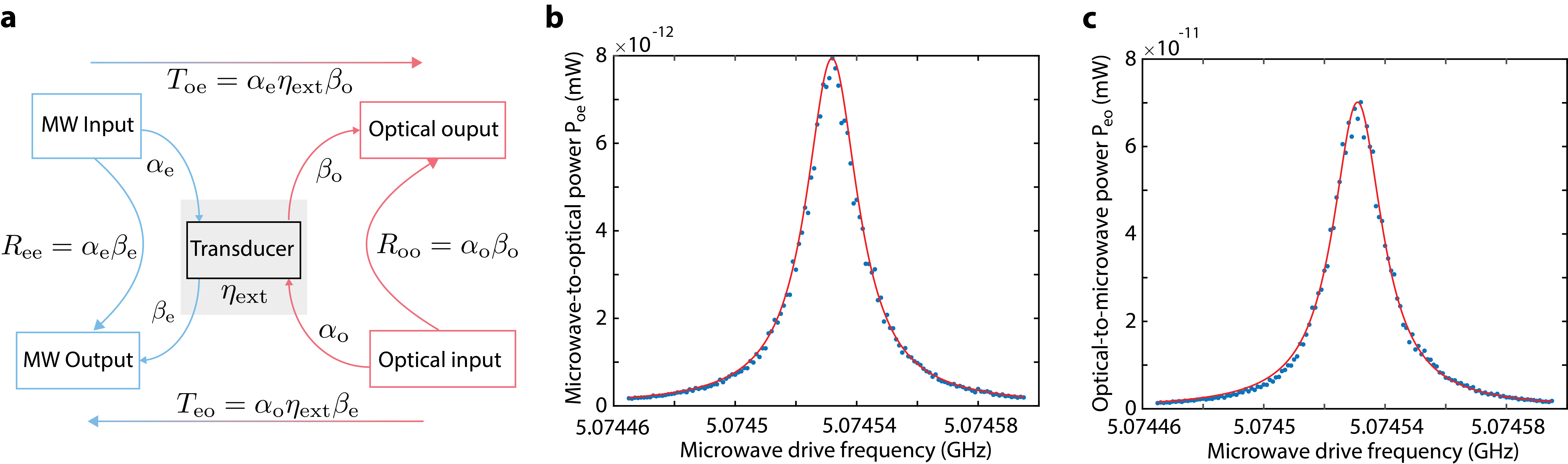}
\caption{\textbf{Characterization of the transduction efficiency.} \textbf{a} Schematic of the four-port measurement. The microwave measurement setup loss (gain) are denoted by $\alpha_{\text{e}}$ ($\beta_{\text{e}}$). Likewise, the optical measurement loss (gain) are given by $\alpha_{\text{o}}$ ($\beta_{\text{o}}$). $\eta_{\text{ext}}$ is the transducer bidirectional microwave-optical transduction efficiency. $T_{\text{oe}}$ ($T_{\text{eo}}$) are the microwave-to-optical (optical-to-microwave) transmittance. $R_{\text{ee}}$ ($R_{\text{oo}}$) are the microwave-to-microwave (optical-to-optical) reflectance. \textbf{b} Measured spectrum of the microwave-to-optical transduction power ($P_{\text{oe}}$) as the microwave input frequency sweeps across the mechanical frequency. \textbf{c} Measured spectrum of the optical-to-microwave transduction power ($P_{\text{eo}}$) as the microwave drive frequency on the optical amplitude modulator sweeps across the mechanical frequency. In \textbf{b} and \textbf{c}, blue dots are measured data whereas red curves are the corresponding Lorentzian fits. We use the peak powers of the fits to calculate the external transduction efficiency in Eq. \ref{eq:4-port}.
}
\label{fig:4-port}
\end{figure}

To measure transduction efficiency, we employ a four-port measurement scheme following Ref. \cite{Andrews2014Apr}, as described below. This method requires four separate transmittance and reflectance measurements between the microwave and optical ports (see \cref{fig:4-port}a).  The measured transmittance in up- and down-conversion can be written in terms of the transducer efficiency
\begin{equation}
T_{\text{oe}}  \stackrel{\text{def}}{=} \frac{P_{\text{oe}}/(\hbar\omega_\text{o})}{P_{\text{e,in}}/(\hbar\omega_\text{e})}   = \alpha_{\text{e}}  \eta_{\text{ext}} \beta_{\text{o}},
\label{eq:P_oe}
\end{equation}
\begin{equation}
T_{\text{eo}}\stackrel{\text{def}}{=}\frac{P_{\text{eo}}/(\hbar\omega_\text{e})}{P_{\text{o,in}}/(\hbar\omega_\text{o})} = \alpha_{\text{o}}  \eta_{\text{ext}} \beta_{\text{e}}.
\label{eq:P_eo}
\end{equation}
Here, $\alpha_{\text{e}}$ ($\alpha_{\text{o}}$) is the power loss from the microwave (optical) input port to the transducer; $\beta_{\text{e}}$ ($\beta_{\text{o}}$) is the power gain from the transducer to the microwave (optical) output port; $\eta_{\text{ext}}$ is the bidirectional conversion efficiency.  

To obtain the transduction efficiency from the transmission coefficients $T_{\text{eo}}$ and $T_{\text{oe}}$, it is necessary to calibrate the microwave/optical input loss and detection gain. To this end, we measure the microwave-to-microwave (optical-to-optical) reflected power $P_{\text{ee}}$ ($P_{\text{oo}}$) at a frequency point that is multiple linewidths away from the microwave (optical) resonator. The reflectance at such a frequency point will not be affected by the transducer response and will only be a function of the gains and losses in the measurement setup
\begin{equation}
R_{\text{ee}}\stackrel{\text{def}}{=} \frac{P_{\text{ee}}/(\hbar\omega_\text{e})}{ P_{\text{e,in}}/(\hbar\omega_\text{e})} =  \alpha_{\text{e}}  \beta_{\text{e}},
\label{eq:P_ee}
\end{equation}
\begin{equation}
R_{\text{oo}}\stackrel{\text{def}}{=} \frac{P_{\text{oo}}/(\hbar\omega_\text{o})}{P_{\text{o,in}} /(\hbar\omega_\text{o})} = \alpha_{\text{o}} \beta_{\text{o}}.
\label{eq:P_oo}
\end{equation}

We can combine the results of these four measurements to find the transduction efficiency as
\begin{equation}
\eta_{\text{ext}} = \sqrt{\frac{T_{\text{eo}}T_{\text{oe}}}{R_{\text{ee}}R_{\text{oo}}}}.
\label{eq:4-port}
\end{equation}
The calculated result from this expression can, in turn, be used to find the product of microwave loss and optical gain of the measurement setup as ($\alpha_{\text{e}}\beta_{\text{o}} = T_{\text{oe}}/\eta_{\text{ext}} $). Using this method, we extract ($\alpha_{\text{e}}\beta_{\text{o}}$) from a set of four-port measurements at a fixed optical power and DC bias. We subsequently measure the transduction efficiency via \cref{eq:P_oe} as we sweep the optical pump power and bias voltage. During these experiments, we periodically repeat the calibration of ($\alpha_{\text{e}}\beta_{\text{o}}$) to ensure we capture slight variations in the optical and microwave gains. 

In measuring microwave-to-optical transmittance ($T_{\text{oe}}$), the input signal at the microwave port is supplied by an RF signal generator. In this experiment, we measure the power of the optical output with a spectrum analyzer in a heterodyne measurement setup (described in \cref{sec:Measurement setup}). The resolution bandwidth for detecting this signal (RBW = 2 kHz) was chosen to be much smaller than the mechanical linewidth while sufficiently large to capture the optical beat note. In measuring optical-to-microwave transmittance ($T_{\text{eo}}$), we generate the input optical signal by modulating the intensity of the optical pump with a microwave tone created by an RF signal generator. In this experiment, we amplify the transducer's microwave output with a chain of amplifiers and detect the output power with a spectrum analyzer. \Cref{fig:4-port}c shows examples of these measurements where the input signal frequency is swept across the mechanical resonance.

 \subsubsection{Effects of mechanical frequency jitter}
\label{sec:Limitations}

As detailed in \cref{sec:TLS}, the mechanical resonator at low optical pump powers experiences frequency jitter. The saturable nature of this effect as a function of microwave drive can create a challenge for precise efficiency characterization. To avoid this challenge, we operate within the range of optical laser powers where this effect is nearly absent. Further, to quantify any residual effects on our measurements, we compare the directly measured microwave-optics transduction efficiency (which requires microwave drives) to the theoretically calculated values using the measured mechanical linewidth in the absence of any microwave drive.

\begin{figure}[htbp]
\centering
\includegraphics[width=0.4\columnwidth]{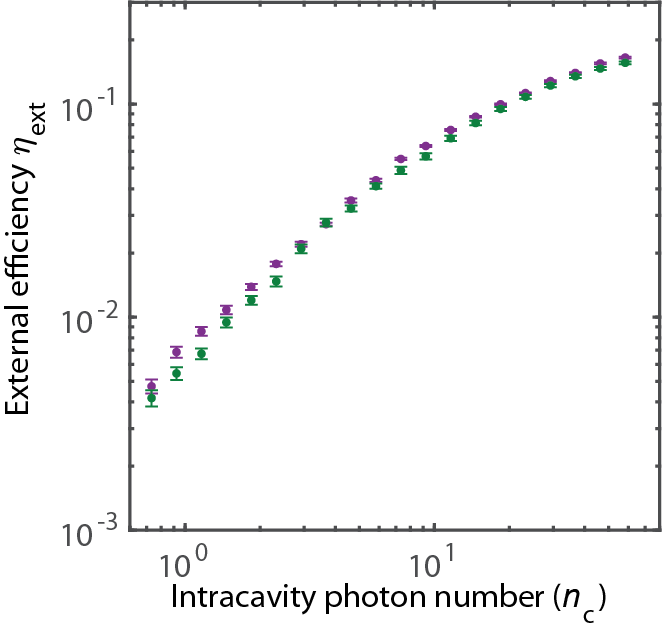}
\caption{\textbf{Comparison of measured and predicted external transduction efficiencies versus intracavity optical photon number at $V_{\text{DC}} = 50\text{ V}$.} Purple data points are the transduction efficiencies obtained via four-port measurements. Green data points are calculated efficiencies using \cref{eq:transduction efficiency} where the total mechanical linewidth is measured without any external microwave drive. Error bars denote 95\% confidence intervals.   
}
\label{figure eff comparison}
\end{figure}

\Cref{figure eff comparison} shows the two data sets. The purple dots depict the direct measurement using the four-port method detailed in \cref{sec:4-port}. In these experiments, we have used an input microwave power of $P_{\text{e,in}}=-58$ dBm.  The green dots in \cref{figure eff comparison} depict the efficiency values calculated using the expression $\eta_{\text{ext}} = \eta_{\text{e}}\eta_{\text{o}}({4\Gamma_{\text{em}}\Gamma_{\text{om}}})/({\Gamma_{\text{tot}}^2}$). For this calculation, we use the measured values of $\Gamma_{\text{em}}$, $\Gamma_{\text{om}}$, $\eta_{\text{o}}$ and $\eta_{\text{e}}$, which are extracted via the methods described in the main text. To find the mechanical linewidth ($\Gamma_{\text{tot}}$), we collect the thermal emission spectrum via the electrical port while setting the microwave input to zero. We observe that for higher optical powers ($n_{\text{c}} \gtrsim 1$), we observe a very good agreement between the two datasets, pointing to near-complete saturation of the frequency jitter by the optical pump in this regime. However, we observe a small difference ($< 22\%$) between the two datasets for the lower end of optical powers in our measurements.

\subsection{Measurement of mechanical and microwave resonator occupancy}
\label{sec:electrical thermometry}
\subsubsection{Electrical thermometry}
\label{sec:Electrical thermometry with gain calibration}

We use electrical thermometry measurements to determine the mechanical and microwave resonator thermal occupancies. In these experiments, a spectrum analyzer is used to measure the symmetrized noise power spectral density $S(\omega)$. It can be shown that this quantity is related to the output microwave field $\hat{a}_{\text{e,out}}$ by \cite{Teufel2011Jul}
\begin{equation}
 \frac{S(\omega)}{\hbar \omega} = \frac{1}{2} \int_{-\infty}^{\infty} d\tau e^{i\omega\tau} \langle \hat{a}_{\text{e,out}}^\dagger(0) \hat{a}_{\text{e,out}}(\tau) + \hat{a}_{\text{e,out}}(\tau)\hat{a}_{\text{e,out}}^\dagger(0) \rangle = \frac{1}{2} \int_{-\infty}^{\infty} d\omega' \langle (\hat{a}_{\text{e,out}}(\omega'))^\dagger \hat{a}_{\text{e,out}}(\omega) + \hat{a}_{\text{e,out}}(\omega)(\hat{a}_{\text{e,out}}(\omega'))^\dagger  \rangle.
\end{equation}

For our devices, we take $n_{\text{o,ext}} \approx n_{\text{o,int}} \approx 0$ due to thermal bath occupancies being negligibly small at optical frequencies. We also take $n_{\text{e,ext}} \approx 0$ since we do not observe any change in the electrical thermal noise background due to the optical pump or microwave drive in any of our experiments. Therefore, the relevant thermal baths for thermometry measurements are $n_{\text{f}}$, $n_{\text{p}}$, and $n_{\text{e,int}}$. By using $\hat{a}_{\text{e,out}}$ derived in \cref{electrical output field}, the symmetrized noise power spectral density becomes  
\begin{equation}
 \frac{S(\omega)}{\hbar \omega} = \frac{1}{2} + \frac{\eta_{\text{e}}\Gamma_{\text{em}}}{(\omega_{\text{m}} - \omega)^2 + (\Gamma_{\text{tot}}/2)^2}(\Gamma_{\text{f}}n_{\text{f}} + \Gamma_{\text{p}}n_{\text{p}})
    +  \frac{(\omega_{\text{m}} - \omega)^2 + \left((\Gamma_{\text{om}} + \Gamma_{\text{i}})/2\right)^2}{(\omega_{\text{m}} - \omega)^2 + (\Gamma_{\text{tot}}/2)^2}|\chi_{\text{e}}(\omega)|^2\kappa_{\text{e,ext}}\kappa_{\text{e,int}}n_{\text{e,int}},
\end{equation}
where $|\chi_{\text{e}}(\omega)|^2 = 1/\left( (\omega_{\text{e}} - \omega)^2 + (\kappa_{\text{e}}/2)^2\right)$  and $\Gamma_{\text{tot}} = \Gamma_{\text{em}} + \Gamma_{\text{om}} + \Gamma_{\text{i}}$. 

In an actual experiment, we obtain the power spectral density after amplifying the microwave field, which requires correcting this expression to account for the gain ($G$, in dB units) and noise ($n_{\text{amp}}$) of the amplification chain \cite{Fink2016Aug}

\begin{equation}
 \frac{S(\omega)}{\hbar \omega} = 10^{G/10}\left( n_{\text{amp}} + \frac{1}{2} + \frac{\eta_{\text{e}}\Gamma_{\text{em}}}{(\omega_{\text{m}} - \omega)^2 + \left(\Gamma_{\text{tot}}/2\right)^2}(\Gamma_{\text{f}}n_{\text{f}} + \Gamma_{\text{p}}n_{\text{p}})
    +  \frac{(\omega_{\text{m}} - \omega)^2 + ((\Gamma_{\text{om}} + \Gamma_{\text{i}})/2)^2}{(\omega_{\text{m}} - \omega)^2 + (\Gamma_{\text{tot}}/2)^2}|\chi_{\text{e}}(\omega)|^2\kappa_{\text{e,ext}}\kappa_{\text{e,int}}n_{\text{e,int}}\right).
\end{equation}

To measure the microwave thermal occupancy, we eliminate the interactions with the mechanical resonator by setting the DC bias to zero, resulting in $\Gamma_{\text{em}} = 0$. Subsequently, the microwave thermal bath occupancy $n_{\text{e,int}}$ can be obtained from the resulting power spectral density
\begin{equation}
\frac{S(\omega)}{\hbar \omega}\Bigr\rvert_{\Gamma_{\text{em}} = 0} = 10^{G/10}\left(n_{\text{amp}} + \frac{1}{2} + \frac{\kappa_{\text{e,ext}}\kappa_{\text{e,int}}}{(\omega_{\text{e}} - \omega)^2 + (\kappa_{\text{e}}/2)^2}n_{\text{e,int}}\right).
    \label{eq:microwave bath occupancy}
\end{equation}
Using the microwave resonator linewidth, the microwave cavity thermal occupancy $n_{\text{mw}}$ is thus determined by \cref{eq: mw occupancy}
\begin{equation}
n_{\text{mw}} = \frac{\kappa_{\text{e,int}}n_{\text{e,int}}}{\kappa_{e}}.
    \label{eq:microwave occupancy}
\end{equation}

The mechanical thermal occupancy can be measured via electromechanical decay into the microwave waveguide set by $\Gamma_{\text{em}}$. In the vicinity of the mechanical mode frequency ($|\omega_{\text{m}} - \omega| << \kappa_{\text{e}}$), we can express the power spectral density as 

\begin{equation}
\begin{split}
 \frac{S(\omega)}{\hbar \omega} &= 10^{G/10}\left( n_{\text{amp}} + \frac{1}{2} + \frac{\eta_{\text{e}}\Gamma_{\text{em}}}{(\omega_{\text{m}} - \omega)^2 + \left(\Gamma_{\text{tot}}/2\right)^2}(\Gamma_{\text{f}}n_{\text{f}} + \Gamma_{\text{p}}n_{\text{p}})
    +  \frac{(\omega_{\text{m}} - \omega)^2 + ((\Gamma_{\text{om}} + \Gamma_{\text{i}})/2)^2}{(\omega_{\text{m}} - \omega)^2 + (\Gamma_{\text{tot}}/2)^2}|\chi_{\text{e}}(\omega)|^2\kappa_{\text{e,ext}}\kappa_{\text{e,int}}n_{\text{e,int}}\right)\\&
    = 10^{G/10}\left( n_{\text{amp}} + \frac{1}{2} + \frac{\eta_{\text{e}}\Gamma_{\text{em}}}{(\omega_{\text{m}} - \omega)^2 + \left(\Gamma_{\text{tot}}/2\right)^2}(\Gamma_{\text{f}}n_{\text{f}} + \Gamma_{\text{p}}n_{\text{p}})
    + \left(1 + \frac{\Gamma_{\text{em}}^2/4 - \Gamma_{\text{em}}\Gamma_{\text{tot}}/2}{(\omega_{\text{m}} - \omega)^2 + (\Gamma_{\text{tot}}/2)^2}\right)|\chi_{\text{e}}(\omega_{\text{m}})|^2\kappa_{\text{e,ext}}\kappa_{\text{e,int}}n_{\text{e,int}}\right).
\end{split}
\end{equation}

For thermometry measurements where the electromechanical cooperativity is purposely made small ($\Gamma_{\text{em}} << \Gamma_{\text{i}}$) by using a large detuning between the microwave and mechanical resonators ($|\omega_{\text{e}} - \omega_{\text{m}}| >> \kappa_{\text{e}}$), the intrinsic mechanical heating rate $\Gamma_{\text{f}}n_{\text{f}} + \Gamma_{\text{p}}n_{\text{p}}$ can be obtained from the resulting power spectral density

\begin{equation}
 \frac{S(\omega)}{\hbar \omega} = 10^{G/10}\left( n_{\text{amp}} + \frac{1}{2} + \frac{\eta_{\text{e}}\Gamma_{\text{em}}}{(\omega_{\text{m}} - \omega)^2 + \left(\Gamma_{\text{tot}}/2\right)^2}(\Gamma_{\text{f}}n_{\text{f}} + \Gamma_{\text{p}}n_{\text{p}})\right).
\end{equation}

For thermometry measurements where the microwave and mechanical resonators are on-resonance ($\omega_{\text{e}} = \omega_{\text{m}}$) and the electromechanical cooperativity is no longer small, the power spectral density becomes 

\begin{equation}
\begin{split}
 \frac{S(\omega)}{\hbar \omega} &= 10^{G/10}\left( n_{\text{amp}} + \frac{1}{2} + \frac{\eta_{\text{e}}\Gamma_{\text{em}}}{(\omega_{\text{m}} - \omega)^2 + \left(\Gamma_{\text{tot}}/2\right)^2}(\Gamma_{\text{f}}n_{\text{f}} + \Gamma_{\text{p}}n_{\text{p}})
    + \left(1 + \frac{\Gamma_{\text{em}}^2/4 - \Gamma_{\text{em}}\Gamma_{\text{tot}}/2}{(\omega_{\text{m}} - \omega)^2 + (\Gamma_{\text{tot}}/2)^2}\right)*4\eta_{\text{e}}\frac{\kappa_{\text{e,int}}n_{\text{e,int}}}{\kappa_{\text{e}}}\right)\\&
    = 10^{G/10}\left( n_{\text{amp}} + \frac{1}{2} + \frac{\eta_{\text{e}}\Gamma_{\text{em}}(\Gamma_{\text{f}}n_{\text{f}} + \Gamma_{\text{p}}n_{\text{p}} + (\Gamma_{\text{em}} - 2\Gamma_{\text{tot}})n_{\text{mw}})}{(\omega_{\text{m}} - \omega)^2 + \left(\Gamma_{\text{tot}}/2\right)^2}
    + 4\eta_{\text{e}}n_{\text{mw}}\right),
\end{split}
\end{equation}
where the effects of noise squashing due to the term containing $n_{\text{e,int}}$ have been fully included in the analysis. 

For both cases of thermometry measurements, the independently measured $n_{\text{mw}}$ from \cref{eq:microwave bath occupancy} can be used to obtain the mechanical mode thermal occupancy by \cref{eq: mechanics occupancy}

\begin{equation}
n_{\text{m}} = \frac{\Gamma_{\text{em}}n_{\text{mw}} + \Gamma_{\text{f}}n_{\text{f}} + \Gamma_{\text{p}}n_{\text{p}}}{\Gamma_{\text{tot}}}.
    \label{eq:mechanical mode occupancy}
\end{equation}

Using \cref{eq:n_add from n_m}, we can also express the input-referred added noise in terms of the mechanical thermal occupancy

\begin{equation}
\begin{split}
    n_{\text{add}} = n_{\text{m}}\frac{\Gamma_{\text{tot}}}{\eta_{\text{e}}\Gamma_{\text{em}}}.
\end{split}
\end{equation}

\subsubsection{Gain calibration by temperature sweep}
\label{sec:electrical gain}

To perform thermometry via the electrical port, we must calibrate the net microwave gain from the transducer device to the spectrum analyzer. This calibration involves measuring the output noise from the amplification chain while varying the temperature of the mixing chamber (MXC) stage in the cryostat ($T_{\text{MXC}}$), where our device is mounted. The measured output noise, integrated over an IF bandwidth $f_{\text{IF}}$ at the spectrum analyzer, comprises amplified thermal noise from the MXC stage and a constant term
\begin{equation}
P = f_{\text{IF}}\hbar \omega 10^{G_\text{A}/10} \left(\frac{1}{e^{\hbar \omega / (k_{\text{B}}T_{\text{MXC}})} - 1} + n_{\text{amp}}\right),
    \label{eq:Gain sweep}
\end{equation}
where $k_{\text{B}}$ is the Boltzmann constant, and $G_{\text{A}}$ is the net gain from the MXC stage to the spectrum analyzer. In this expression, the constant term $n_{\text{amp}}$ denotes the total added noise of the amplifier chain, which is dominated by the contribution from a high electron mobility transistor (HEMT) amplifier located at the 4K stage of the cryostat.

\begin{figure}[htbp]
\centering
\includegraphics[width=0.5\columnwidth]{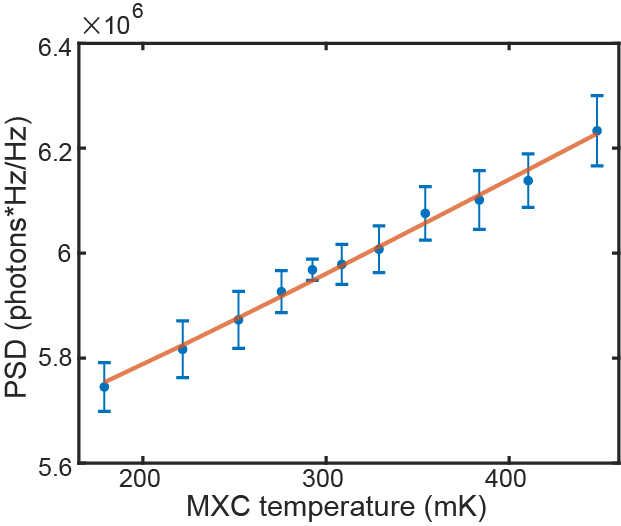}
\caption{\textbf{Detected power spectral density vs MXC temperature.} Error bars denote the measured power spectral density with 95\% confidence interval. Red line is the fit from which the electrical gain $G_{\text{A}}$ is obtained.  
}
\label{figure rf gain}
\end{figure}

We control the MXC temperature by reducing the $^3$He/$^4$He mixture flow and applying a heater with varying power at the MXC stage. This method keeps the HEMT amplifier's temperature nearly constant. Consequently, any observed changes in noise with temperature are attributed to thermal noise from the MXC, which is multiplied by the net gain $G_\text{A}$ before being detected. \Cref{figure rf gain} shows the measured noise power spectral density versus the MXC stage temperature, from which we extract $G_{\text{A}} = 56.59 \pm 0.29$ dB.

Despite the relatively small statistical uncertainty in this value, this method of calibrating the gain suffers from systematic error due to its inability to account for any loss at the MXC temperature that happens between the device and the HEMT amplifier. As an alternative, we employ an optical method for calibrating the electrical gain, which is free from this systematic error and detailed in the following section.

\subsubsection{Gain calibration by optical sideband asymmetry}
\label{sec:Optical sideband asymmetry}

In our devices, optical sideband asymmetry provides a calibration-free means of determining mechanical thermal occupancy. The mechanical thermal occupation found from this method can, in turn, be used as a calibrated source for the electromechanical readout to provide the net microwave gain $G$ as detailed below.

In the low optomechanical cooperativity limit ( i.e., so long as the back-action does not influence $n_\text{m}$), the difference between the red- and blue-sideband scattering rates corresponds to the vacuum contribution (that is, spontaneous Stokes scattering), which can serve as a reference for finding the mechanical thermal occupation as \cite{Underwood2015Dec,Meenehan2015}
\begin{equation}
\frac{1}{n_{\text{m}}} = \frac{\int_{-\infty}^{+\infty} S_{\text{b}}(\omega) \,d\omega}{\int_{-\infty}^{+\infty} S_{\text{r}}(\omega) \,d\omega}-1.
\label{eq:sideband asymmetry n_m}
\end{equation}
Here, $S_{\text{r}}(\omega)$ and $S_{\text{b}}(\omega)$ represent the noise power spectral density of the mechanical mode under the red- and blue-sideband pumps, respectively.

\begin{figure}[htbp]
\centering
\includegraphics[width=0.5\columnwidth]{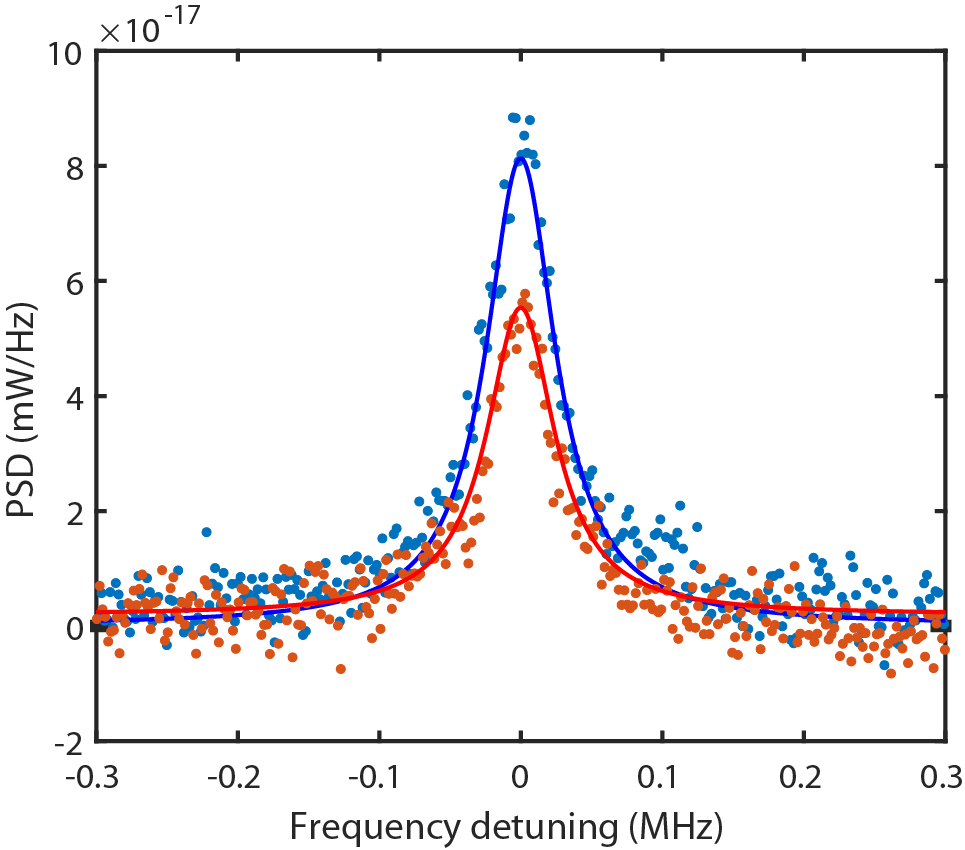}
\caption{\textbf{Spectra of mechanical mode noise power density under red- and blue-sideband pumps.} In these measurements, we use $n_{\text{c}} = 9.2$ and $V_{\text{DC}} = 40$ V, corresponding to an optomechanical cooperativity of $C_\text{om} \approx 0.04$. The noise floor corresponds to $16.4$ noise photons (referred to the optical waveguide at the chip) by the optical heterodyne detection setup and is removed from the plots. Solid curves are Lorentzian fits to the corresponding data. To eliminate any uncertainty from wavelength-dependent optical waveguide transmission \cite{Weaver2024Feb}, we ensure the optomechanical back-action rate remains equal for these two measurements by monitoring the microwave-to-optical transduction efficiency and the mechanical linewidth.
}
\label{figure sideband asymmetry}
\end{figure}

\Cref{figure sideband asymmetry} shows the optically measured power spectral density of the mechanical motion for blue- and red-pumps for the same intracavity photon number. Using \cref{eq:sideband asymmetry n_m}, we calculate the mechanical thermal occupancy for these measurements as $n_{\text{m}} = 1.81 \pm 0.35$. We subsequently measure the mechanical thermal emission via the microwave port under the same optical pump conditions. The microwave power detected in this measurement is then referred to the mechanical occupation from the optical sideband asymmetry ($n_{\text{m}} = 1.81 \pm 0.35$) to find the microwave output line gain as $G = 55.83 \pm 0.84$ dB. We have used this calibrated gain in all the measurements reported in the main text.

\subsection{Optical measurements of input-referred added-noise}
\label{sec:optical noise}
The input-referred added noise of microwave-to-optical transduction is defined as the number of microwave photons at the input port of the transducer that would result in the same optical noise measured at the transducer optical output. A straightforward approach of characterizing the input-referred added noise is to optically measure the transducer noise and normalize it to the input microwave photon number. In our experiments, we use optical heterodyne detection, as described in \cref{sec:Measurement setup}, to measure the transducer's output optical noise. Under a red-detuned optical pump without any external microwave input, the optical noise photon quanta at the photodetector is given by 
\begin{equation}
N_{\text{o,noise}} = 10^{G_{\text{o}}/10}n_{\text{o,out}}(\omega_{\text{m}})  = \frac{P_{\text{o,noise}}} {f_{\text{IF}} \hbar \omega_{\text{o}}}, 
    \label{eq:optical noise}
\end{equation}
where $G_{\text{o}} = 7.9 \text{ dB} $ is the optical detection gain,  $n_{\text{o,out}}(\omega_{\text{m}})$ is the optical noise spectral density at the waveguide (see \cref{optical spectral density}), and $P_{\text{o,noise}}$ is the detected optical noise power at the mechanical resonator frequency within an IF bandwidth $f_{\text{IF}}$ = 2 kHz. Due to the low optomechanical readout rate ($C_{\text{om}} << 1$), measuring the optical noise power via continuous heterodyne detection took $\sim 17$ hours for the lowest intracavity photon number reported in the main text ($n_{\text{c}} = 5.8$) and becomes impractically long for lower photon numbers.  

To obtain the input-referred added noise, we normalize $N_{\text{o,noise}}$ to an equivalent microwave photon number at the transducer microwave waveguide. For this purpose, we measure the total conversion efficiency at the mechanical resonator frequency from the transducer microwave waveguide to the photodetector, which can be measured as 
\begin{equation}
\eta_{\text{tot}} = 10^{G_{\text{o}}/10} \eta_{\text{ext}}(\omega_{\text{m}}) = \frac{P_{\text{oe}}/\left(f_{\text{IF}} \hbar \omega_{\text{o}}\right)} {P_{\text{e}}/\left(f_{\text{IF}} \hbar \omega_{\text{e}}\right)}, 
    \label{eq:photon conversion}
\end{equation}
where $\eta_{\text{ext}}$ is the transducer conversion efficiency between the on-chip microwave and optical waveguides, $P_{\text{e}}$ is the input microwave power at the transducer microwave waveguide, and $P_{\text{oe}}$ is the detected power of the coherent transducer output optical signal. Dividing the detected optical noise by $\eta_{\text{tot}}$, we obtain the input-referred added noise as
\begin{equation}
n_{\text{add}} = N_{\text{o,noise}} / \eta_{\text{tot}} = n_{\text{o,out}}(\omega_{\text{m}})/\eta_{\text{ext}}(\omega_{\text{m}}). 
    \label{eq:photon conversion}
\end{equation}

\section{Performance comparison}
\begin{table}[h!]
\begin{tabular}{|c|c|c|}
\hline
 \bf{Device} & $\mathbf{n_{\text{\bf{add}}}}$ & $\mathbf{\eta_{\text{\bf{ext}}} \times B \times D}$ \bf{(Hz)} \\ 
  \hline
  This work & 0.94 & 1900 \\ 
  \hline
  This work & 0.58 & 470 \\ 
  \hline
 Weaver et al. Nat Nanotech (2024) & 6.2 & 5.2 \\   
 \hline
 Brubaker et al. PRX (2022) & 3.2 & 130 \\  
 \hline
  Jiang et al. Nat Phys (2023) & 1.6 & 1100 \\
 \hline
 Mirhosseini et al. Nature (2020) & 0.57 & 0.075 \\
 \hline
 Sahu et al. Nat Comm (2022) & 0.4 & 5.4 \\
 \hline
 Meesala et al. Nat Phys (2024) & 0.14 & 3.1 \\
 \hline
\end{tabular}
\caption{Transducer Comparison}
\label{transducer comparison}
\end{table}

To compare different transducers with each other (\cref{transducer comparison}), we consider two figures of merit: 1) input-referred added noise $n_{\text{add}}$, where $n_{\text{add}} < 1$ defines the quantum-enabled regime \cite{Zeuthen2020May, Sahu2022Mar, Kumar2023Mar}, and 2) the product of external efficiency $\eta_{\text{ext}}$, bandwidth $B$, and duty cycle $D = T_{\text{d}} \times R_{\text{p}}$, where $T_{\text{d}}$ = pulse duration and $R_{\text{p}}$ = pulse repetition rate. 
 
 In Ref. \cite{brubaker_optomechanical_2022}, the lowest achieved added noise is $n_{\text{add}} = 3.2$ with a reported external efficiency of $\eta_{\text{ext}} =0.47$ and a bandwidth of $B = 220 \text{ Hz}$. To separate out the device properties from those of the characterization setup, we make the conservative choice of excluding the effects of the reported coupling loss from spatial mode mismatch in this experiment and use $\eta_{\text{ext}} = 0.59$ for calculations. This transducer operates using continuous pumps, so $D = 1$. In Ref. \cite{Sahu2022Mar}, the highest efficiency operation point in the quantum-enabled regime is reported with the added noise of $n_{\text{add}} = 0.4$ and efficiency of $\eta_{\text{ext}} = 0.15$.
To separate out the device properties from those of the characterization setup, we make the conservative choice of excluding the effects of the reported coupling loss from spatial mode mismatch in this experiment and use $\eta_{\text{ext}} = 0.25$ for calculations. In this experiment, the transducer bandwidth is reported as $B = 18 \text{ MHz}$ and the pulse duration is $T_{\text{d}} = 300$ ns with a repetition rate of $R_{\text{p}} = 4$ Hz. The reported efficiency in \cite{Weaver2024Feb,Mirhosseini2020Dec,Jiang2023Oct,Meesala2024Feb}  is defined for a temporal mode matching the pulse shape, which equals $\eta_{\text{ext,p}} = \eta_\text{ext}\times B\times T_{\text{d}}$. In Ref. \cite{Weaver2024Feb}, the lowest added noise and the external photon conversion efficiency are reported as $n_{\text{add}} = 6.2$ and $\eta_{\text{ext,p}} = 5.2 \times 10^{-5}$. The pulse repetition rate for this experiment is $R_{\text{p}} = 100 \text{ kHz}$. In Ref. \cite{Mirhosseini2020Dec}, the lowest achieved added noise is $n_{\text{add}} = 0.57$ with $\eta_{\text{ext,p}}= 7.5 \times 10^{-4}$. The pulse repetition rate is $R_{\text{p}} = 100 \text{ Hz}$. In Ref. \cite{Jiang2023Oct}, the added noise in the microwave waveguide is $n_{\text{add}} = 1.6$ with $\eta_{\text{ext,p}}= 6.3 \times 10^{-3}$, which is calculated by the product of the optical photon-phonon scattering probability ($0.036$), the phonon-microwave photon conversion efficiency ($0.35$), optical external coupling efficiency ($0.5$), and microwave external coupling efficiency ($0.993$). The pulse repetition rate is $R_{\text{p}} = 170 \text{ kHz}$. In Ref. \cite{Meesala2024Feb}, the added noise in the microwave waveguide is $n_{\text{add}} = 0.14$, which is calculated as $\eta_{\text{mw}}\times n_{\text{i}}+n_{\text{d}}$ according to the numbers reported in its supplement section 13. The external conversion efficiency at the low noise operation point is $\eta_{\text{ext,p}} = 6.1 \times 10^{-5}$, which is calculated by the product of the optical photon-microwave photon scattering probability ($1.8 \times 10^{-4}$), optical external coupling efficiency ($0.5$), and microwave external coupling efficiency ($0.68$). The pulse repetition rate is reported as $R_{\text{p}} = 50 \text{ kHz}$.

\end{document}